\documentclass[reprint,aps,prl,showpacs,twocolumn]{revtex4-1}

\usepackage[utf8]{inputenc}
\usepackage{soul}
\usepackage{mathptmx}
\usepackage{mathtools}
\usepackage{amsmath} 
\usepackage{braket,ulem}
\usepackage{float}
\DeclarePairedDelimiter\abs{\lvert}{\rvert}%
\usepackage{lipsum}  
\usepackage{svg}
\let\vec\mathbf
\usepackage{mathptmx}
\usepackage{stackrel}
\usepackage{mathtools}
\usepackage{amsmath} 
\usepackage{braket}
\usepackage{float}
\usepackage[colorlinks= true, linkcolor=blue, citecolor=blue, urlcolor=blue]{hyperref}
\usepackage{lipsum}  
\usepackage{svg}
\let\vec\mathbf
\renewcommand\thesection{\arabic{section}}
\usepackage[sort&compress]{natbib}
\usepackage{mhchem}
\usepackage{changepage}
\usepackage{float}
\usepackage{mathrsfs}
\usepackage{tikz}
\usepackage{amsmath}
\usepackage{mathtools}
\usepackage{mathrsfs}
\usepackage{esint}
\usepackage{amssymb}
\usepackage{eucal}
\usepackage{tikz}

\date{}

\begin{document}

\title{Microscopic Theory of Adsorption Kinetics}

\author{{\normalsize{}Yuval Scher$^{1,}$}
{\normalsize{}}}\email{yuvalscher@mail.tau.ac.il}

\author{{\normalsize{}Ofek Lauber Bonomo$^{1}$}
{\normalsize{}}}

\author{{\normalsize{}Arnab Pal$^{2,3}$}
{\normalsize{}}}\email{arnabpal@imsc.res.in}

\author{{\normalsize{}Shlomi Reuveni$^{1,}$}
{\normalsize{}}}\email{shlomire@tauex.tau.ac.il}

\affiliation{\noindent \textit{$^{1}$School of Chemistry, Center for the Physics \& Chemistry of Living Systems, Ratner Institute for Single Molecule Chemistry, and the Sackler Center for Computational Molecular \& Materials Science, Tel Aviv University, 6997801, Tel Aviv, Israel}}

\affiliation{\noindent \textit{$^{2}$The Institute of Mathematical Sciences, CIT Campus, Taramani, Chennai 600113, India}}

\affiliation{\noindent \textit{$^{3}$Homi Bhabha National Institute, Training School Complex, Anushakti Nagar, Mumbai 400094, India}}

\date{\today}


\begin{abstract}
Adsorption is the accumulation of a solute at an interface that is formed between a solution and an additional gas, liquid, or solid phase. The macroscopic theory of adsorption dates back more than a century and is now well-established. Yet, despite recent advancements, a detailed and self-contained theory of \textit{single-particle adsorption} is still lacking. Here, we bridge this gap by developing a microscopic theory of adsorption kinetics, from which the macroscopic properties follow directly. One of our central achievements is the derivation of the microsopic version of the seminal Ward-Tordai relation, which connects the surface and subsurface adsorbate concentrations via a universal equation that holds for arbitrary adsorption dynamics. Furthermore, we present a microscopic interpretation of the Ward-Tordai relation which, in turn, allows us to generalize it to arbitrary dimension, geometry and initial conditions. The power of our approach is showcased on a set of hitherto unsolved adsorption problems to which we present exact analytical solutions. The framework developed herein sheds fresh light on the fundamentals of adsorption kinetics, which opens new research avenues in surface science with applications to artificial and biological sensing and to the design of nano-scale devices.
\end{abstract}

\maketitle

The field of adsorption kinetics is over a century old, dating back to the seminal works of Langmuir \cite{langmuir1916constitution,langmuir1917constitution,langmuir1918adsorption}. Back then, it was less than a decade since Perrin -- based on Einstein's theory of diffusion -- provided concrete experimental evidence for the atomic nature of matter  \cite{perrin2013brownian,einstein1905molekularkinetischen}. It is thus hardly surprising that the brilliant scientists of that era focused on macroscopic models of adsorption -- it is very likely that most did not foresee single-particle and nano scale experiments becoming the tangible reality of our days \cite{elenko2009single,elenko2010single,van2011single,van2014single,gong2014direct,shen2016single,wang2017three,borberg2019light,wang2019probing,wang2020probing,wang2020non}. Recent decades have seen a constant improvement in single-particle technology and experimentation \cite{Betzig2016,Hell1994,Moerner1997,Xie1998,Block1999,Bustamante2003,Gaub1997}. However, a systematic microscopic kinetic theory of adsorption, as far as we know, is still lacking. Such theory is required since macro- and microscopic kinetics differ not only in the mathematical approach to the modeling of the problem, but also in the fundamental questions that they allow one to pose and answer. To bridge this gap, we now set to construct a detailed theory of single-particle adsorption kinetics. The potential uses of this theory and its wide applicability are illustrated by examples and discussion. The remainder of this paper is structured as follows.

\begin{figure}[t!]
\begin{centering}
\includegraphics[width=1\linewidth]{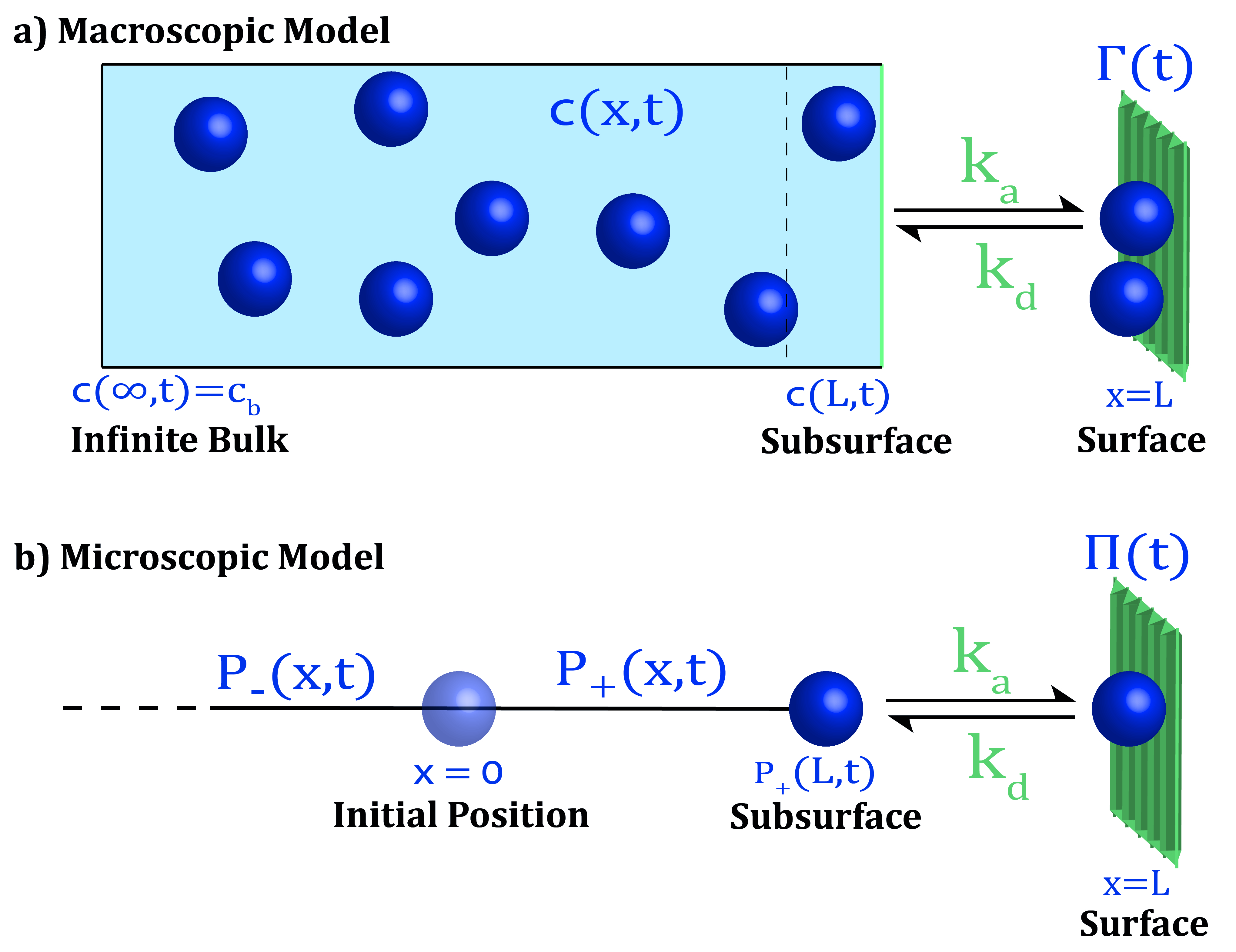}
\caption{a) Illustration of the most common macroscopic adsorption model. A surface of area $\mathcal{A}$, here located at $x=L$, is immersed in an infinite solution of initial uniform concentration $c_{b}$. The surface concentration, $\Gamma(t)$, is initially zero, and its value at later times depends on the adsorption dynamics at the subsurface layer $c(L,t)$. Usually, linear adsorption and desorption rates are assumed, with $k_{a}$ and $k_{d}$ taken to be the adsorption and desorption rate constants respectively. In the Langmuirian model, a maximal surface concentration that cannot be exceeded is assumed, and the adsorption rate is proportional to the surface fraction which is bare.
b) The analogous microscopic adsorption model. Here $\Pi(t)$ is the probability to find the particle on the surface at time $t$, and $P_{\pm}(x,t)$ is the probability density to find it in the bulk to the left/right of the initial position $x=0$.}
\end{centering}
\end{figure}

In Sec. I we provide a brief review of the theory of macroscopic adsorption kinetics, from equilibrium to dynamic adsorption. The vast majority of theoretical works were conducted for the 1D model of a surface immersed in an infinite bulk, with initial uniform concentration, as illustrated in Fig. 1a. The surface concentration depends on the transport of adsorbates to the surface (e.g., via diffusion) and the adsorption kinetics at the subsurface layer. In the review section we survey fundamental results that were obtained by assuming various types of adsorption kinetics. We give special attention to the Ward-Tordai relation, which is a general relation between the surface and subsurface concentrations that holds for unspecified adsorption dynamics.

In Sec. II we develop the analogous theory for single-particle adsorption, as illustrated in Fig. 1b. Naturally, single-particle models require us to shift to a probabilistic perspective -- for example, instead of the surface concentration, we are interested in the probability to find the particle on the surface. Furthermore, we derive the microscopic version of the Ward-Tordai relation and show that the macroscopic version can be derived from it. We believe that the physical basis that explains why the Ward-Tordai relation takes its form was not adequately accounted for previously. We thus give a modern interpretation of this relation for both macro and micro adsorption dynamics.

To conclude Sec. II we exemplify the theory of single-particle adsorption by solving two fundamental adsorption models. In Sec. II.C we solve for an adsorbate diffusing in effective 1-dimensional semi-infinite tube (Fig. 1b). This problem is the analogue of the classical macroscopic adsorption model (Fig. 1a). In Sec. II.D we consider the case of an adsorbate that is diffusing in channel,  effectively 1-dimensional, that is confined between two adsorbing surfaces. We also attain the solution for the analogous macroscopic problem, with uniform initial concentration (assuming a dilute enough solution such that the adsorbates do not interact with one another). Even this relatively simple macroscopic model was considered only once before \cite{adamczyk1987adsorption}, and we believe that the treatment presented here is clearer.

In Sec. III we draw insight from the observations made in Sec. II and generalize the theory for any dimension, geometry and adsorption kinetics. Specifically, in Sec. III.A we generalize the Ward-Tordai relation beyond its 1-dimensional version. Also, in Sec. III.B, we  demonstrate how to apply the theory in high dimensions by solving explicitly for a model system where an adsorbate diffuses in a spherical shell, i.e., between two concentric spheres. By taking the radius of the inner sphere to zero, or the radius of the outer sphere to infinity, we obtain as special cases results for two systems that are of particular interest: diffusion of an adsorbate inside and outside an adsorbing sphere, respectively.

In Sec. IV we briefly recapitulate the results obtained in this paper, and discuss them in relation to sister problems from other disciplines which share some similarities. We hope that this discussion will supply ideas for further research.

\section{I. Brief Review of the Theory of Macroscopic Adsorption Kinetics}

\subsection{A. Equilibirum Adsorption (Isotherms)}

When an interface is formed between a solution and any other phase (be it gas, liquid or solid), adsorption is the accumulation of the solute at the surface phase \cite{chattoraj1984miscellaneous, baret1968kinetics}. Here we will assume that the surface is well defined and of width of a few molecular layers. In macroscopic adsorption studies, the main interest is usually modeling the equilibrium adsorption isotherm, namely, a curve depicting the surface concentration ($[\text{mol}\cdot m^{-2}]$) at equilibrium, $\Gamma_{eq}$, as function of bulk pressure (for gas adsorbates) or concentration/activity (for liquid adsorbates), where the temperature is kept constant \cite{foo2010insights}. Empirical isotherm models date back to the Freundlich equation, and are of use when theory based models fail \cite{adamson1967physical}. Here, however, we will focus on isotherms that were derived from basic principles. The simplest isotherm one can conceive is a linear adsorption isotherm, also known as the Henry isotherm, in which $\Gamma_{eq}$ is proportional to the bulk  pressure/concentration (note the resemblance to Henry's law) \cite{chang1995adsorption}. It takes only a simple experiment to show that the linear model naturally breaks at high pressures/concentrations, as the surface gets saturated with adsorbates. 

It was Langmuir who proposed the first complete kinetic model that accounts for the saturation effect \cite{langmuir1916constitution,langmuir1918adsorption}. In his model he assumed that: i) the desorption rate from a second layer is rapid enough to allow for a single-layer model; ii) the adsorbates do not interact with each other; and iii) the adsorption rate is homogeneous across the surface and proportional to the fraction of surface that is bare, $1-\Gamma_{eq}/\Gamma_{\mathrm{m}}$, where $\Gamma_{m}$ is the maximal surface concentration. By equating the adsorption term to a linear desorption term he got an expression for $\Gamma_{eq}=\frac{K c}{1+Kc}\Gamma_{m}$, where $K$ is the equilibrium constant, and $c$ is the adsorbate bulk concentration. 

Langmuir's model has proved extremely efficient when coming to understand adsorption isotherms, and it is still widely used these days. Numerous generalizations and alternative models were proposed over the years \cite{foo2010insights}. Notable are these which proposed remedies to cases in which Langmuir's model assumptions are invalid. Examples are the BET model which treats multi-layer formation \cite{brunauer1938adsorption}, and the chemisorption models of Ehrlich and Kisliuk which consider heterogeneous surfaces \cite{ehrlich1956mechanism,kisliuk1957sticking}. Bulk or surface hydrolysis, protein unfolding/denaturation and aggregation of adsorbates in the solution at high concentrations can complicate the problem even further \cite{chang1995adsorption, noskov2020adsorption}.

\subsection{B. Dynamic Adsorption}

In the late 30's a large amount of experimental data accumulated on time variations in surface tension in liquid-air interfaces before equilibrium is established \cite{ward1946time}. In the late 30's and early 40's few researchers theorized that this dynamic surface tension is diffusion-influenced \cite{bond1937lxxvii,langmuir1937effect,ross1945change}. The work of Ward and Tordai is usually regarded as the first comprehensive treatment of the role of diffusion in adsorption processes \cite{ward1946time}. Ward and Tordai borrowed from Carslaw's treatment of heat conduction in solids \cite{carslaw1921introduction} to model the concentration on the surface, here taken to be located at $L$, as emanating from two contributions: i) diffusion from the bulk and ii) back-diffusion from the surface:

\begin{equation} \label{WardTordai}
  \Gamma(t)=\sqrt{\frac{D}{\pi}}\left[2 c_{\mathrm{b}} \sqrt{\mathrm{t}}-\int_{0}^{t} \frac{c(L,\tau)}{\sqrt{t-\tau}} d \tau\right],  
\end{equation}

\noindent where $c_{\mathrm{b}}$ is the bulk initial uniform concentration ($\lim_{x \to -\infty} c(x,t) = c(x,0)=c_{b}$), $D$ is the adsorbate diffusivity and $c(L,t)$ is the adsorbate concentration at the subsurface layer. It is interesting to note that, after numerical comparison of data and theory, Ward and Tordai concluded that, for the systems they have studied, it is unlikely that diffusion is a rate-determining process, and one should consider an additional activation barrier \cite{ward1946time, ward1952time}. Nonetheless they have paved the way to further theoretical work on diffusion-influenced dynamic surface tension. 

The difficulty in solving Eq. (\ref{WardTordai}) is due to the unknown subsurface concentration $c(L,t)$. Essentially, Eq. (\ref{WardTordai}) relates between the surface concentration and the concentration at the subsurface, but an additional relation is required for solving. This additional information is not given by the Ward-Tordai model, and should instead be introduced as an external boundary condition \cite{diamant1996kinetics}. One approach is to assume that the surface and subsurface are in quasi-equilibrium, such that the equilibrium relation between $c(L,t)$ and $\Gamma(t)$ is also valid out of equilibrium. For instance, assuming that $\Gamma(t)=K c(L,t)$ for all times (the assumption of the Henry isotherm). This case was first solved by Sutherland \cite{sutherland1952kinetics} and later by Delahay and Trachtenberg \cite{delahay1957adsorption} and by Hansen \cite{hansen1961diffusion}. Note, however, that in assuming this relation, one introduces externally the kinetics that takes place on the boundary, which in this case is equivalent to the introduction of an activation barrier. 

In solving, the authors of Ref. \cite{sutherland1952kinetics,delahay1957adsorption,hansen1961diffusion}  used a straightforward approach, different from Ward and Tordai, of defining exactly what is the boundary value problem to be solved. Like Ward and Tordai, they have assumed that the bulk has an initial uniform concentration, while the surface is initially empty:

\begin{subequations}  \label{Hansen1}
\begin{align}
      c(x,0) = c_{b}, \\
     \Gamma(0)=0,        
\end{align}
\end{subequations}

\noindent that the bulk is so vast that it can be safely assumed that the concentration far from the boundary is unaffected by the adsorption to the surface

\begin{equation} \label{Hansen2}
    c(x \to -\infty ,t) = c_{b},
\end{equation}

\noindent and that transport in the bulk follows the diffusion equation

\begin{equation} \label{Hansen3}
   \frac{\partial c(x,t)}{\partial t}=D\frac{\partial^2 c(x,t)}{\partial x^2}.
\end{equation}

\noindent They have postulated the mass-balance condition 

\begin{equation} \label{Hansen4}
   \frac{d\Gamma (t)}{dt} = -D\frac{\partial c(x,t)}{\partial x} \Big|_{x=L} =J(L,t),
\end{equation}

\noindent where $J(L,t)$ is the diffusive flux to the surface. Finally, they supplied an additional closure relation that defines the boundary condition:

\begin{equation} \label{Hansen5}
\Gamma(t)=Kc(L,t).
\end{equation}

\noindent As discussed above, the relation in Eq. (\ref{Hansen5}) assumes local quasi-equilibrium between the surface and the subsurface. 

 In the aforementioned works the authors proceeded to Laplace transform Eqs. (\ref{Hansen1}-\ref{Hansen5}) and solved the resulting linear system of equations to get an explicit expression for $\tilde{\Gamma}(s)$. After inverse Laplace transformation the authors reported:
 
\begin{equation} \label{Hansen6}
 \frac{\Gamma(t)}{\Gamma_{eq}} = 1 - e^{\frac{D t}{K^2}} \text{erfc} \left(\frac{\sqrt{D t}}{K}\right),
\end{equation}

\noindent where $\Gamma_{eq}=K c_{b}$ is the equilibrium surface concentration and $\text{erfc}(\cdot)$ is the complementary error function \cite{hansen1961diffusion,delahay1957adsorption,chang1995adsorption}.

Markedly, by Laplace transforming Eqs. (\ref{Hansen1}-\ref{Hansen4}) only, and omitting the closure relation of Eq. (\ref{Hansen5}), Hansen obtained a linear system of equations, that after some algebra gives a relation between $\tilde{\Gamma}(s)$ and $\tilde{c}(L,s)$

\begin{equation} \label{WardTordaiLaplace}
  \tilde{\Gamma}(s) =  \frac{\sqrt{D}}{s^{3/2}}c_{b}-\sqrt{\frac{D}{s}}\tilde{c}(L,s),
\end{equation}

\noindent which upon inverse Laplace transformation yields the Ward-Tordai relation of Eq. (\ref{WardTordai}). In fact, in this work we will use this method to derive the analogous single-particle relation. 

It is clear that the Henry isotherm (Eq. (\ref{Hansen5})) is valid only in the case of dilute solutions, and should break down at higher concentrations as the surface gets saturated. Delahay et. al. considered the case in which the closure relation is given by the Langmuir isotherm instead of the Henry isotherm \cite{delahay1957adsorption,delahay1958adsorption}. Although the Langmuir isotherm is more suitable for macroscopic dynamics, the resulting system of equations is non-linear and analytical solutions to it are unknown. Researches then resorted to numerical methods \cite{miller1981solution} and to a regular perturbation series method \cite{mccoy1983analytical}. 

Apart from considering different isotherms as closure relations, other modifications for the scheme above were suggested. For instance, one can consider different geometries \cite{mysels1982diffusion,frisch1983diffusion,adamczyk1987adsorption}, or extend the Ward-Tordai relation to include micellar solutions \cite{liu2009diffusion,miller1991adsorption}, evaporation effects \cite{hansen1960theory}, ionic surfactants \cite{miller1991adsorption} and more.

Without a closure relation that permits an analytical solution of Eq. ($\ref{WardTordai}$), its application to experimental data must be performed numerically, while approximating the convolution integral \cite{hansen1960theory,miller1991adsorption,nguyen2006effect,ziller1986solution,li2010simple}. For example, following Miller and Kretzschmar \cite{miller1991adsorption}, it is convenient to rewrite Eq. ($\ref{WardTordai}$) in a different form

\begin{equation} \label{WardTordaiForNumerics}
  \Gamma(t)=2\sqrt{\frac{D}{\pi}}\left[ c_{\mathrm{b}} \sqrt{\mathrm{t}}-\int_{0}^{\sqrt{t}} c(L,t-\tau) d \sqrt{\tau}\right],  
\end{equation}

 \noindent where we have used the convolution property $\int_{0}^{t} \frac{c(L,\tau)}{\sqrt{t-\tau}} d \tau = \int_{0}^{t} \frac{c(L,t-\tau)}{\sqrt{\tau}} d \tau$, and changed variables $\tau \to \sqrt{\tau}$. Given a closure relation, for example the Langmuir isotherm relation, Eq. (\ref{WardTordaiForNumerics}) can be integrated numerically.

\subsection{C. Diffusion-kinetic (mixed-kinetic) adsorption}
 
By using the adsorption isotherms for the closure relation of the aforementioned boundary problem one actually assumes diffusion-controlled adsorption. Namely, that the timescale for equilibration of the surface and subsurface is very fast compared to the diffusion timescale \cite{fernandez2011numerical}. The introduction of the full diffusion-kinetic model is due to Baret \cite{baret1968kinetics,liggieri1996diffusion}. A general treatment of diffusion-kinetics adsorption was proposed by Borwankar and Wasan \cite{borwankar1983kinetics}. In their work, a general kinetic relation on the surface boundary was defined:

\begin{equation} \label{Borwankar1}
    \frac{d \Gamma(t)}{d t}=r_{1}(t)-r_{-1}(t),
\end{equation}

\noindent where $r_{1}(t)$ and $r_{-1}(t)$ are the adsorption (forward ) and desoprtion (backward) terms of the kinetic expression. Borwankar and Wasan went further and postulated that it is generally true that only the adsorption term is dependent on subsurface concentration $c(L,t)$, and that this dependence is linear. Furthermore, both terms depend on the surface concentration. Equation (\ref{Borwankar1}) can then be re-written as:

\begin{equation} \label{Borwankar2}
\frac{d \Gamma(t)}{d t}=k_{a} G\left(\Gamma(t)\right) c(L, t)-k_{d} H(\Gamma(t)),
\end{equation}

\noindent where $k_{a}$ and $k_{d}$ are the adsorption and desorption rates, and $G(\Gamma(t))$ and $H(\Gamma(t))$ are some general functions of the surface concentration. Here we have defined Eq. (\ref{Borwankar2}) in a slightly different manner than in the original paper, but this is a matter of notation only. An expression for $c(L, t)$ can be extracted from Eq. (\ref{Borwankar2}) and plugged into  Eq. (\ref{WardTordai}), yielding the Ward-Tordai relation for the case of diffusion-kinetic adsorption. 

Two main kinetic expressions are usually considered. The first is that of linear kinetics

\begin{equation} \label{Henry}
 \frac{d \Gamma(t)}{d t} = k_{a} c(L,t) - k_{d} \Gamma(t).
\end{equation}

\noindent It can be easily seen that at equilibrium, i.e., when $ \frac{d \Gamma(t)}{d t} = 0 $, one gets the Henry isotherm with $K = k_{a}/k_{d}$.

The second is consistent with the Langmuir isotherm \cite{chang1995adsorption}:

\begin{equation} \label{Langmuir}
    \frac{d \Gamma(t)}{d t}=k_{a} c(L,t)\left(1-\frac{\Gamma(t)}{\Gamma_{\mathrm{m}}}\right)-k_{d} \Gamma(t),
\end{equation}

\noindent where $\Gamma_{m}$ is the maximal surface concentration, and by setting $ \frac{d \Gamma(t)}{d t} = 0 $ one gets the Langmuir isotherm $\Gamma_{eq}=\frac{K c}{1+Kc}\Gamma_{m}$ with $K = \frac{k_{a}}{k_{d}\Gamma_{m}}$.

A solution for Eqs. (\ref{Hansen1}-\ref{Hansen4}) together with Eq. (\ref{Henry}) as a closure relation was given by Adamczyk (Eq. 16 in Ref. \cite{adamczyk1987nonequilibrium}). Here again, using the Langmuirian relation of Eq. (\ref{Langmuir}) as the closure relation leads to a non-linear integro-differential equation, and so Miura and Seki employed numerical methods and upper/lower boundary estimations \cite{miura2015diffusion}.

\section{II. Theory for Single-Particle Adsorption}


\subsection{A. The Microscopic Ward-Tordai Relation}
We will now derive a microscopic analogue of Eq. (\ref{WardTordai}). Recall that this equation relates between the surface concentration $\Gamma(t)$ and the subsurface concentration $c(L,t)$. When we treat a single-particle, however, we move to a probabilistic view -- we now aim to find a general relation between the probability to be on the surface, and the probability to be at the subsurface. The second difference is of course the initial conditions. Equation (\ref{WardTordai}) holds for a 1-dimensional semi-infinite line where initially there is a uniform concentration of adsorbates that spreads out to infinity. For the single-particle analogue, we consider a particle that is initially at a certain distance from the surface with probability one, thus having zero probability to be found elsewhere. A precise formulation of the problem is given below.

Consider an adsorbate, initially at $x=0$, diffusing on a 1-dimensional semi-infinite line, with an adsorbing surface located at $x=L>0$ (Fig. 1b). At the subsurface, the particle is adsorbed with a constant rate $k_{a}$. Once adsorbed, it desorbs with a constant rate $k_{d}$. The probability of being on the surface at time $t$ is denoted by $\Pi(t)$. The diffusion propagator which gives the probability density to find the particle in the bulk is denoted $p(x,t)$ and divided by two restrictions, $p_{\pm}(x,t)$, where the $\pm$ determines whether the particle is right (+) or left (-) of its initial position  $x=0$.

We consider an ordinary diffusion for which the propagator obeys

\begin{equation} \label{Micro1}
  \frac{\partial p(x,t)}{\partial t} =  D\frac{\partial^2 p(x,t)}{\partial x^2}, 
\end{equation}

\noindent with the following initial conditions

\begin{subequations}  \label{Micro2}
\begin{align}
    & p(x,0) = \delta(x),   \\
   &  \Pi(0)=0,        
\end{align}
\end{subequations}

\noindent and the boundary condition

\begin{equation}  \label{Micro3}
     p_{-}(x \to -\infty,t) = 0.
\end{equation}

\noindent Equations (\ref{Micro2}) and (\ref{Micro3}) are the single-particle analogues of Eqs.  (\ref{Hansen1}) and (\ref{Hansen2}), respectively.

We retain the mass-balance condition (Eq. (\ref{Hansen4})) by simply replacing  $c(x,t)$ with $p(x,t)$ and $\Gamma(t)$ with $\Pi(t)$, and noting that for $x>0$ we have  $p(x,t) = p_{+}(x,t)$:

\begin{equation} \label{Micro4}
  \frac{d\Pi (t)}{dt} =  -D\frac{\partial p_{+}(x,t)}{\partial x} \Big|_{x=L}.
\end{equation}

Note that in modeling the single-particle case we have moved to a probabilistic description, hence normalization must hold

\begin{equation}  \label{Micro5}
     \Pi(t) + \int_{-\infty}^{L} p(x,t)dx = 1.
\end{equation}

\noindent In fact, Eq. (\ref{Micro5}) is equivalent to Eq. (\ref{Micro4}) and the two can be used interchangeably. To see this, take the derivative of Eq. (\ref{Micro5}) with respect to time and then use the diffusion equation. 

Similarly to Hansen's approach in Ref. \cite{hansen1961diffusion}, we Laplace transform Eqs. (\ref{Micro1}-\ref{Micro4}) and solve for $\tilde{\Pi}(s)$ to obtain (Appendix A)

\begin{equation} \label{MicroWardTordaiLaplace}
  \tilde{\Pi}(s) =  \frac{e^{-L\sqrt{\frac{s}{D}}}}{s}-\sqrt{\frac{D}{s}}\tilde{p}(L,s),
\end{equation}

\noindent which differs from Eq. (\ref{WardTordaiLaplace}) only in the first term on the right-hand side. Upon inverse Laplace transformation Eq. (\ref{MicroWardTordaiLaplace}) yields:

\begin{equation} \label{MicroWardTordai}
  \Pi(t)=\text{erfc} \left(\frac{L}{2 \sqrt{ D t}}\right)-\sqrt{\frac{D}{\pi}}\int_{0}^{t} \frac{p(L,\tau)}{\sqrt{t-\tau}} d \tau,  
\end{equation}

\noindent which is the single-particle analogue of the Ward-Tordai relation of Eq. (\ref{WardTordai}). In the next subsection we will see how in fact  Eq. (\ref{WardTordai}) can be obtained from  Eq. (\ref{MicroWardTordai}).

In the derivation of Eq. (\ref{MicroWardTordai}) we have assumed that the particle is initially in the bulk at a distance L from the surface. Let us consider the case were the particle starts on the surface. In this case, there is no need to divide the propagator using restrictions, and Eq. (\ref{Micro2}) is modified to

\begin{subequations}\label{Micro6} 
\begin{align} 
    & p(x,0):=p_0(x) = 0,   \\
   &  \Pi(0)=1.        
\end{align}
\end{subequations}

\noindent Solving (Appendix A), one gets 

\begin{equation} \label{MicroWardTordaiLaplaceStartSurface}
  \tilde{\Pi}(s) = \frac{1}{s} -\sqrt{\frac{D}{s}}\tilde{p}(L,s),
\end{equation}

\noindent which differs from of Eq. (\ref{MicroWardTordaiLaplace}) only in the first term on the right-hand side. Inverse Laplace transforming Eq. (\ref{MicroWardTordaiLaplaceStartSurface}) yields

\begin{equation} \label{MicroWardTordaiStartSurface}
  \Pi(t)=1-\sqrt{\frac{D}{\pi}}\int_{0}^{t} \frac{p(L,\tau)}{\sqrt{t-\tau}} d \tau.  
\end{equation}

Finally, note that for both models considered above, we also solved for the Laplace transform of the propagator $\tilde{p}(x,s)$ in terms of $\tilde{\Pi}(s)$, or alternatively in terms of $\tilde{p}(L,s)$. These expressions and their derivations for all initial positions are given in Appendix A. When the particle starts on the surface the result is especially compact and is given by 

\begin{equation} \label{SurfaceBulkRelationStartSurface}
  \tilde{p}(x,s) = \sqrt{\frac{s}{D}} e^{(x-L) \sqrt{\frac{s}{D}}} \left[\frac{1}{s}-\tilde{\Pi}(s)\right] ,
\end{equation}

\noindent where $x<L$. 

\subsection{B. A modern interpretation of the Ward-Tordai relation}
In hindsight, it is perhaps not surprising that the macroscopic Ward-Tordai relation (Eq. (\ref{WardTordai})) and its microscopic analogue (Eq. (\ref{MicroWardTordai})) are similar in form, and differ only in the first term. 

In their paper, Ward and Tordai explain the first term in Eq. (\ref{WardTordai}) as a term corresponding to the diffusive flow of adsorbates from the bulk region (which is initially at concentration $c_b$) to the surface (initially at zero concentration). The second term, they say, represents the flow of adsorbates in the opposite direction, from the surface to the bulk \cite{ward1946time}. We feel that this explanation, that prevailed in later works on adsorption dynamics, is rather vague and perhaps a tad anachronistic. Hence, we would like to offer here a modern interpretation of the relation. 

Ward and Tordai borrowed relevant results from Carslaw's book on the mathematical theory of heat conduction \cite{carslaw1921introduction}. In here, our discussions rely upon the results in \cite{carslaw1921introduction} and \cite{carslaw1959conduction} which was published after Ward and Tordai's seminal work.

\subsubsection{First term - diffusive flow of adsorbates from the bulk}

Starting from the first term on the right-hand side of Eq. (\ref{WardTordai}), what did Ward and Tordai meant by ``the diffusive flow of adsorbates from the bulk region"? The first term actually gives the concentration of adsorbates accumulated on the surface by time $t$, given the initial condition, and given that every adsorbate arriving at the surface adsorbs immediately and stays adsorbed. This is exactly the problem of diffusion to an \textit{absorbing} boundary (note difference between the ad- and ab- prefixes), but instead of the particles being annihilated on the boundary, they are just counted as adsorbed. Namely, while the absorbing boundary does not conserve probability, the adsorbing boundary does, as we keep counting the adsorbed particles as part of the system. Let us denote this concentration of irreversibly adsorbed particles by $M(t)$. For the initial condition considered by Ward and Tordai, that of uniform initial concentration that spreads to infinity, it is a simple task to show that \cite{carslaw1959conduction}

\begin{equation} \label{wardM}
    M(t) =c_{\mathrm{b}} \sqrt{\frac{4 D t}{\pi}},
\end{equation}

\noindent which is indeed the first term of Eq. (\ref{WardTordai}). This expression was first considered by Langmuir and Schaefer \cite{langmuir1937effect}. 

When considering a single-particle we move again to a probabilistic description. The quantity that we are after is simply the cumulative distribution of the so called first-passage distribution \cite{redner2001guide,klafter2011first,metzler2014first}. The first-passage density, which we will denote by $f(t)$, is the probability density function that a particle arrives at the \textit{absorbing} surface for the first time at a certain time $t$, given the initial condition. Thus, its cumulative distribution, $F(t)=\int_{0}^{t} f(t')\text{d}t'$, gives the probability that a particle arrived at the surface by time $t$, and was irreversibly adsorbed.  

For the delta-function inital condition considered in Eq. (\ref{Micro2}) we indeed have $F(t)=\text{erfc} \left(\frac{L}{2 \sqrt{ D t}}\right)$ \cite{redner2001guide}. Of course, one can think of any other normalized initial condition and find the proper cumulative distribution. However, note that in this case $F(t)$ will simply be given by an integration of $\text{erfc} \left(\frac{L}{2 \sqrt{ D t}}\right)$ over the initial distribution. In fact, we can calculate in this manner $M(t)$ of unnormalizable concentration profiles as well.  For example, if we assume an initial uniform concentration $c_{b}$, we have 

\begin{equation}
    M(t) = \int_{0}^{\infty} c_{b} \text{erfc} \left(\frac{L}{2 \sqrt{ D t}}\right) dL = c_{b}\sqrt{\frac{4 D t}{\pi}},
\end{equation}

\noindent which is indeed the expression in Eq. (\ref{wardM}).

Thus, we can express the first term in the Ward-Tordai relation for any initial condition and surface dynamics in terms of the solution to the corresponding absorbing boundary problem with delta-function initial condition. This is not a mere coincidence -- the delta function initial condition is special, its propagator is the so called Green's function.  From here onwards we will denote the Green's function of the absorbing boundary problem by $g(x,t \mid x',\tau)$, namely the probability to find a diffusing single-particle at $x$ at time $t$, given that it was introduced to the semi-infinite line at $x'<L$ at time $\tau \leq t$, and in the presence of an absorbing boundary at $x=L$. In what follows, we will show that the Green's function can be utilized to (re)derive, interpret, and generalize the Ward-Tordai relation.

The calculation of the Green's function for the Ward-Tordai model turns out to be particularly simple and elegant by use of the method of images \cite{redner2001guide}. We know that a particle that diffuses freely on the 1-dimensional infinite line has a Gaussian propagator. Explicitly, the probability density to find the particle at $x$ at time $t$ given that the particle started at $x'$ at time $\tau$ is given by $\frac{1}{\sqrt{4 \pi D (t-\tau)}} e^{-\frac{(x-x')^2}{4 D (t-\tau)}}$. Imagine now that the initial position of our particle is $x'<L$ and that there is another freely diffusing \textit{anti-particle} starting at the same time at $2L-x'$ (a mirror image with respect to the absorbing boundary which is located at $L$). Now, we assume that whenever the particle and anti-particle meet, they annihilate and instantly disappear. Mathematically we take this into account by putting a minus sign in front of the anti-particle Gaussian density. Note that the anti-Gaussian is also a solution of the diffusion equation, and so the linear combination of the particle and anti-particle densities is by itself a solution of the diffusion equation: 

\begin{equation} \label{GreensFunction}
    g(x,t \mid x',\tau) = \frac{ e^{-\frac{(x-x')^2}{4 D (t-\tau)}} - e^{-\frac{(x-2L+x')^2}{4 D (t-\tau)}}}{\sqrt{4 \pi D (t-\tau)}} .
\end{equation}

\noindent Now, note that by plugging $x'=L$ we have $p(x,t \mid x'=L,\tau)=0$, i.e., this linear combination is zero on the boundary, which is exactly the meaning of an absorbing boundary condition. Hence, by simply ignoring the imaginary construction at $x>L$, we see that Eq. (\ref{GreensFunction}) is the solution for the model of an adsorbate diffusing on the semi-infinite line $-\infty<x<L$ with an absorbing boundary at $L$. Thus, Eq. (\ref{GreensFunction}) gives the Green's function required for the solution of the Ward-Tordai model.

Finally, note that the first-passage distribution to an absorbing surface, given an initial condition $p_0(x')$, is simply the diffusive flux of the Green's function into the surface,

\begin{equation} \label{firstpassage}
   f(t) = -D \int_{-\infty}^{L} p_0(x') \frac{\partial g(x,t \mid x',\tau) }{ \partial  x} \Big|_{x=L} dx',
\end{equation}

\noindent and recall that $F(t)=\int_{0}^{t} f(t')\text{d}t'$. Thus, we can express the first-term of the Ward-Tordai relation in terms of the Green's function, and more precisely in terms the derivative of the Green's function with respect to the position coordinate $x$.

\subsubsection{Second term - flow of adsorbates from the surface to the bulk}

We have seen that the first term of the Ward-Tordai relation gives the concentration on the surface in the limit of immediate adsorption and no desorption. Of course, the whole idea of the model is to allow for adsorption kinetics on the surface, and indeed the second term of the Ward-Tordai is a correction term that accounts for exactly that. Some of the particles that arrive at the subsurface are not immediately adsorbed, and of course some of the adsorbed particles desorbs back into the subsurface. The exact amount of these adsorbates is determined by an external boundary condition that specifies the subsurface concentration $c(L,t)$. Out of these adsorbates, some may end up adsorbed to the surface at time $t$, and we do not need to correct for them. The rest of the adsorbates will end up somewhere in the bulk, and these are the ones we need to correct for. We do so by subtracting their amount from the first term in the Ward-Tordai relation.

As we did in the analysis of the first term, let us first consider the single-particle picture. We want to use again the Green's function, this time with $x'=L$, to give the probability to start at the subsurface at time $\tau$ and be found at some point $x$ in the bulk after some additional time $t-\tau$. Unfortunately, by the definition of the Green's function the probability density at $x'=L$ is zero at all times (the Green's function is for a case where the boundary is absorbing, such that starting on it means instantaneous absorption). We thus need to be more careful. 

Let us first examine the  propagator for a case where the adsorbate starts at a very small distance from the surface. Plugging $x' = L - \Delta x$ in Eq. (\ref{GreensFunction}) and expanding the exponents to first order in $\Delta x$ we obtain:

\begin{equation} \label{GreensFunctionExpansion}
    g(x,t \mid L-\Delta x,\tau) \simeq  \frac{L-x}{2\sqrt{\pi} \left [D (t-\tau)\right]^{3/2}} e^{-\frac{(L-x)^{2}}{4 D(t-\tau)}}\Delta x~.
\end{equation}

\noindent If we now take the limit $\Delta x \to 0$ we get $ g(x,t \mid L-\Delta x,\tau) \to 0$, which brings us back to square one. However, a finite probability density at the subsurface requires $g(x,t \mid L-\Delta x,\tau) \to C(\tau)$, where $C(\tau)$ is some function that corresponds to the required subsurface probability density $p(L,\tau)$. To achieve this we counterbalance the loss of density as $\Delta x$ is made small by increasing the magnitude of the injected density at time $\tau$ by a factor $\epsilon(\tau)$ such that $\lim_{\Delta x \to 0}(\epsilon \Delta x)=C(\tau)$, and   

\begin{align} \label{LimitDoublet}
  \lim_{\Delta x\to 0} \epsilon(\tau) g(x,t \mid L- & \Delta x,\tau) = 
  \\
 & C(\tau) \frac{L-x}{2\sqrt{\pi} \left [D (t-\tau)\right]^{3/2}} e^{-\frac{(L-x)^{2}}{4 D(t-\tau)}}. \nonumber
\end{align}

\noindent Comparing with the Taylor expansion in Eq. (\ref{GreensFunctionExpansion}), we note that one can re-write Eq. (\ref{LimitDoublet}) as 

\begin{equation}  \label{Doublet}
  \lim_{\Delta x \to 0} \epsilon(\tau) g(x,t \mid L-\Delta x,\tau) = - C(\tau)\frac{\partial g(x,t \mid x',\tau) }{ \partial x'} \Big|_{x'= L}. 
\end{equation}

In the procedure above we have merely multiplied a solution of the diffusion equation (Eq. (\ref{GreensFunction})) by a constant, and took the limit $\Delta x \to 0$, where higher orders of the expansion do not contribute, and the approximation in Eq. (\ref{GreensFunctionExpansion}) is exact. Therefore, Eq. (\ref{Doublet}) is by itself a solution of the diffusion equation with an absorbing surface. A solution of this form is called a doublet (or a dipole in electrostatics).

The external boundary condition specifies $p(L,\tau)$ for all $t> \tau$. To find the probability density at $x$ at time $t$ due the corrective density on the surface we time integrate over a doublet with time-dependent magnitude $C(\tau)=D p(L,\tau)$

\begin{equation} \label{Convolution}
\begin{array}{ll}
- D \int_{0}^{t} p(L,\tau)\frac{\partial g(x,t \mid x',\tau) }{ \partial x'} \Big|_{x'=L} d \tau,
\end{array}
\end{equation}

\noindent where in Appendix B we show why a doublet of this magnitude keeps a subsurface concentration of $p(L,\tau)$ for all $\tau<t$. Note that Eq. (\ref{Convolution}) is in the form of a convolution.

Finally, by spatial integration of Eq. (\ref{Convolution}) over the bulk, we obtain the density that was introduced at the subsurface due to the external boundary condition and is still in the bulk by time $t$,

\begin{align} \label{Correction}
 -D\int_{-\infty}^{L}\int_{0}^{t} p(L,\tau)\frac{\partial g(x,t \mid x',\tau) }{ \partial x'} \Big|_{x'=L} d \tau & d x  =
 \\
 &\sqrt{\frac{D}{\pi}}\int_{0}^{t} \frac{p(L,\tau)}{\sqrt{t-\tau}} d \tau. \nonumber
\end{align}

\noindent This is exactly the probability that needs to be subtracted from the first term of the Ward-Tordai relation to correct for the neglected surface dynamics. Thus, the second term in the Ward-Tordai relation can be expressed in terms of the Green's function derivative as well, this time with respect to the initial position $x'$, where we recall that in the first term the derivative is with respect to $x$. 

Furthermore, we can utilize the fact that the cumulative distribution below Eq. (\ref{firstpassage}) can be written as $F(t) = 1-Q(t)$, where $Q(t)$ is the survival probability, namely, the probability to find the adsorbate somewhere in the bulk by time t, given that the surface is absorbing. For the 1-dimensional scenario we have $Q(t)=\int_{-\infty}^{L} \left[\int_{-\infty}^{L} g(x,t \mid x',0)  p_{0}(x') dx' \right] dx$. Thus, the first term can be equivalently written in terms of the Green's function itself.

Recapitulating, the 1-dimensional Ward-Tordai relation generalized to arbitrary initial condition $p_{0}(x)$ can be expressed in terms of the Green's function as follows

\begin{equation}  \label{WardTordaiInTermsOfGreens}
\begin{array}{ll} 
\Pi(t) = \underbrace{1-\int_{-\infty}^{L} \left[\int_{-\infty}^{L} g(x,t \mid x',0)  p_{0}(x') dx' \right] dx}_{\textrm{First term, positive}}
\\
\\
\underbrace{+D\int_{-\infty}^{L}\int_{0}^{t} p(L,\tau)\frac{\partial g(x,t \mid x',\tau) }{ \partial x'} \Big|_{x'=L} d \tau d x}_{\textrm{Second term, negative}}.
\end{array}
\end{equation} 

Note that in order to obtain the macroscopic version of Eq. (\ref{WardTordaiInTermsOfGreens}), we simply replace the probability $\Pi(t)$ and the probability densities $p(L,\tau)$ and $p_{0}(x)$ with the concentrations  $\Gamma(t)$, $c(L,\tau)$ and $c_{0}(x)$, respectively, and the $1$ in the first term with $N/\mathcal{A}$, where $N$ is the overall number of adsorbates and $\mathcal{A}$ is the surface area. Then, the second term is the surface concentration that needs to be subtracted from the first term to correct for the neglected surface dynamics.

\subsection{C. Diffusion in a semi-infinite tube with an adsorbing surface}

Let us consider the single-particle adsorption problem described in Sec. II.A and depicted in Fig. 1b. Specifically, we assume the validity of Eqs. (\ref{Micro1})-(\ref{Micro4}). Let us also assume the linear kinetics of Eq. (\ref{Henry}) as the closure relation.
By Laplace transforming this system of equations and solving, we obtain (Appendix C)

 \begin{equation}  \label{surface}
\tilde{\Pi}(s) = \frac{ K e^{-L\sqrt{\frac{s}{D}}}}{ K s + (\frac{s}{k_{d} }+1)\sqrt{D s}},
 \end{equation}

\noindent where  $K:=k_{a}/k_{d}$ is the equilibrium constant for the adsorption-desorption process (See Sec. I.C). 

In Sec. IV we identify and elaborate on the mapping between adsorption problems and Smoluchowski-type reversible (association-dissociation) reactions in 1-dimension. The corresponding microscopic 1-dimensional dissociation-recombination problem was solved by Agmon in Ref. \cite{agmon1984diffusion}. Under the correct re-scaling, one minus Eq. (28) therein gives the inversion of Eq. (\ref{surface}) above. To show this, we analytically invert Eq. (\ref{surface}) in appendix D, and explain the steps involved in detail. The resulting form is quite bulky and so is not presented in the main text. Instead, we  explore the result in different limits.

There are two ways in which $K \to 0$: either $k_{d} \to \infty$ or $k_{a} \to 0$. In both cases the surface is completely non-adsorbing, and is thus akin to a reflecting surface. Indeed, in this limit the right-hand side of Eq. (\ref{surface}) goes to zero.
In the limit of $k_{a} \to 0$, the particle cannot be adsorbed from the subsurface. Since, by the initial condition given in Eq. (\ref{Micro2}.b), we have $\Pi(0)=0$, the probability $\Pi(t)$ vanishes identically for all $t>0$. This limit can also be understood by revisiting Eq. (\ref{Henry}). In the microscopic analogue, the adsorption term is zero because $k_{a} \to 0$ and the desorption term is zero because $\Pi(t)=0$. We thus get $J(L,t)=0$, which is exactly the reflecting boundary condition. Similarly, in the limit of $k_{d} \to \infty$ the desorption is immediate, and so effectively the particle spends no time on the surface.

In the limit $k_{a} \to \infty$ we get $\tilde{\Pi}(s) = \frac{e^{-L\sqrt{\frac{s}{D}}} }{s}$. In this limit, whenever the particle desorbs back to the subsurface it is instantly re-adsorbed. Effectively, once the particle reaches the surface it is stuck there. Hence, the surface acts as a pseudo-absorbing boundary. The only difference is that the system still conserves probability, since the particle is counted even when adsorbed. Inverting $\tilde{\Pi}(s)$ in this limit we get $\Pi(t)=\text{erfc}(\frac{L}{2\sqrt{Dt}})$ \cite{spiegel1965schaum}, which is exactly the cumulative first-passage probability---from the origin to an absorbing boundary at $L$---of a particle diffusing on a semi-infinite line. Thus, in this limit, the probability to be at the boundary gradually saturates at 1. 

In fact, we have already dealt and interpreted this function in Sec.II.B. In the context of that discussion, note that the second term in the microscopic Ward-Tordai equation (Eq. (\ref{MicroWardTordai})) goes to zero after plugging in the closure relation in Eq. (\ref{Henry}) and taking the limit $k_{a} \to \infty$.  This makes perfect sense, as this term provides correction for imperfect adsorption or desorption which are negligible in the limit $k_{a} \to \infty$. Note, however, that this is not true when $k_{d} \to \infty$ simultaneously, in such a way that $k_{a}$ and $k_{d}$ are comparable. For instance, in the limit $k_{a}=k_{d} \to \infty$, the Laplace transform of the probability to be on surface is given by $\tilde{\Pi}(s) = \frac{  e^{-L\sqrt{\frac{s}{D}}}}{  s + \sqrt{D s}}$. This result can be inverted to give $\Pi(t) = e^{L+Dt}\text{erfc}(\frac{L+2Dt}{2\sqrt{Dt}})$, which can be used as an elegant approximation for the probability to be on the surface when $k_{a}$ and $k_{d}$ are large and equal.

By taking the limit $k_{d} \to 0$ we kill the desorption term in Eq. (\ref{Henry}), which becomes a ``radiating" boundary condition \cite{collins1949diffusion,sano1979partially,weiss1986overview,szabo1984localized,singer2008partially,palreactive,grebenkov2020imperfect}, where the range between the limiting reflecting and absorbing boundary conditions is spanned by tuning $k_{a}$ from zero to infinity. Note, however, that in contrast to the classical radiating boundary condition here probability is conserved overall by accounting for the probability to be found on the boundary.    

\begin{figure}[t]
\begin{centering}
\includegraphics[width=1\linewidth]{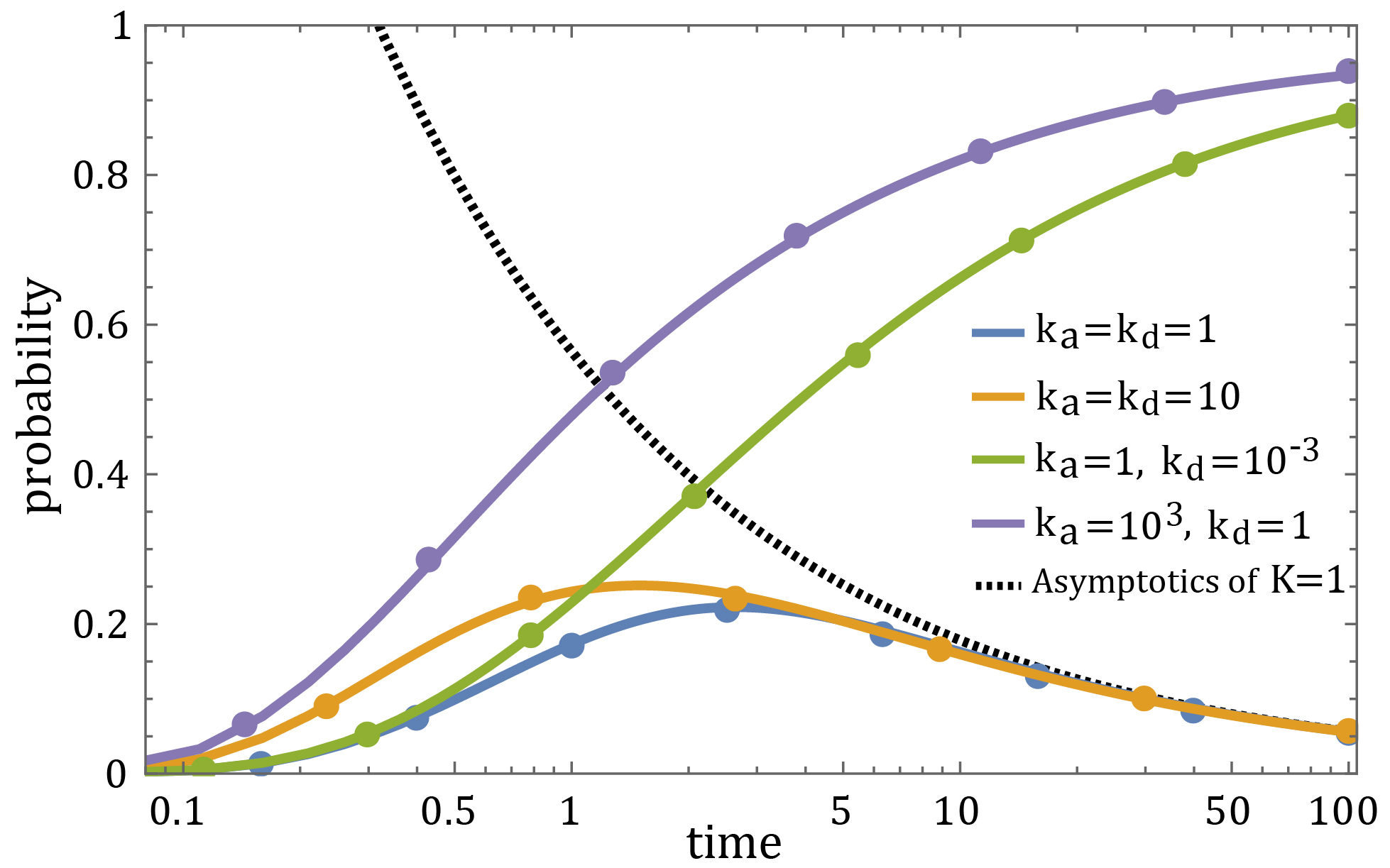}
\caption{A log-linear plot of the probability to find the particle on the adsorbing surface vs. time. The adsorbate diffuses on the semi-infinite line and adsorption kinetics are linear. Full circles come from simulations, and the solid lines are plotted by inverting Eq. (\ref{surface}). Here we have taken $L=1$ and $D=1$ throughout. The blue line is for the case of $k_a = k_d =1$, and the orange line is for the case $k_a = k_d =10$. Both these curves share the same equilibrium constant $K=k_a/k_d=1$. Consequently they also share the long time asymptotics of Eq. (\ref{surface2}), which is depicted by the dashed black curve. The green and purple curves represent choices of very low $k_d$ and very high $k_a$, thus mimicking ``absorbing" and ``radiating" surfaces in the observation time under consideration. These also follow the large time asymptotics of Eq. (\ref{surface2}) (not shown in plot).}
\end{centering}
\end{figure}

Of course, we must also consider the possibility that both $k_{d}$ and $k_{a}$ are finite and comparable. We can gain additional understanding of the result in Eq. (\ref{surface}) by considering the long time asymptotics, corresponding to the limit $s \to 0$. By expanding Eq. (\ref{surface}) around $s=0$ and inverse Laplace transforming the first term of the expansion, we obtain the following long-time asymptotics

\begin{equation}  \label{surface2}
\Pi(t) \simeq \frac{K}{\sqrt{\pi Dt}}.
\end{equation}

\noindent Note that, if $D$ is known, Eq. (\ref{surface2}) implies that $K$ can be easily extracted from this aymptotic behaviour. Also note that the same behaviour was observed for 1-dimensional dissociation-recombination reactions in Eq. (3.4) of Ref. \cite{agmon1989theory} (more on that in Sec. IV).

In Fig. 2 we plot the results of Monte-Carlo simulations for various values of $k_{a}$ and $k_{d}$. In all cases, there is a critical time $t^{*}$ in which $\Pi(t)$ attains its maximum, which is followed by a decay that has the asymptotics of Eq. (\ref{surface2}). The critical time $t^{*}$ can be determined numerically. Knowing $t^{*}$ may be useful when constructing an experiment, since at this time the probability to find the particle on the surface is the highest.

Now consider a variation of this problem, where the particle is initially on the boundary itself (for the corresponding dissociation-association problem and inversion see Appendix C of Ref. \cite{agmon1988geminate}). Equation (\ref{Micro2}) is then replaced by Eq. (\ref{Micro6}) and there is no need to divide $p(x,t)$ using restrictions. 

In Appendix C, we show that the solution to this problem is given by

\begin{equation}  \label{surface3}
\tilde{\Pi}(s) =  \frac{K + \frac{\sqrt{D s}}{k_d} }{ K s + (\frac{s}{k_{d} }+1)\sqrt{D s}},
\end{equation}
and
\begin{equation}  \label{bulk1}
\tilde{p}(x,s) =  \frac{1}{ K s + (\frac{s}{k_{d} }+1)\sqrt{D s} }  e^{(x-L)\sqrt{\frac{s}{D}}}.
\end{equation}

\noindent  It can be easily seen that Eqs. (\ref{surface3}) and  Eqs. (\ref{bulk1}) are indeed related via Eq. (\ref{SurfaceBulkRelationStartSurface}).

In the limit $K \to \infty$ we get $\Pi(t) = 1$. This is expected, since in this limit the particle is sure to stay adsorbed to the surface. In the limit $k_{a} \to 0$ we get $\Pi(t) = e^{-k_{d} t}$, i.e., the particle desorbs with rate $k_{d}$ and never re-absorbs. In the limit $k_{d} \to \infty$ we get $\Pi(t) = 0$, i.e., the desorption is instantaneous and afterwards the surface is simply reflective. Not surprisingly, the long time asymptotics are independent of the initial position. Thus, Eq. (\ref{surface2}) is also valid here.

Finally, we define $\Lambda(t)$ as the ensemble averaged duration of the time spent on the surface in the course of the observation time $t$

\begin{equation}  \label{name}
\Lambda(t) = \int_0^t \Pi(t') dt',
\end{equation}

\noindent such that $\tilde{\Lambda}(s) = \frac{1}{s} \tilde{\Pi}(s)$. Thus, by multiplying Eqs. (\ref{surface}) and (\ref{surface3}) by $1/s$ we get $\tilde{\Lambda}(s)$ when starting in the bulk and on the surface respectively. While the corresponding inversions $\Lambda(t)$ are as difficult to compute as $\Pi(t)$, the long time asymptotics is simply given by

\begin{equation}  \label{name2}
\Lambda(t) \simeq \frac{2 K \sqrt{t}}{\sqrt{\pi D}}.
\end{equation}

\subsection{D. Diffusion in a tube confined between two adsorbing surfaces}

Consider an adsorbate, initially at $-L<x_{0}<L$, diffusing in an interval of length $2L$ with two identical adsorbing surfaces at $-L$ and at $L$. Let $\Pi_{\pm}(t)$ be the probabilities to find the particle on the right (+) and left (-) surfaces at time $t$. We define $\Pi(t) = \Pi_{+}(t) + \Pi_{-}(t)$, the probability to find the particle on either of the surfaces at time $t$. The initial condition for this problem thus reads

\begin{subequations}\label{box1}  
\begin{align} 
   &  p(x,0) = \delta(x-x_0),   \\
  &   \Pi(0)=\Pi_{\pm}(0)=0.        
\end{align}
\end{subequations}

Recall that the probability density to find the particle in the bulk is $p_{\pm}(x,t)$, where $\pm$ discriminates positions that are to the right (+) or left (-) of the initial position $x_0$. The boundary conditions for the right and left surfaces are given by a combination of Eq. (\ref{Micro4}) and the linear-kinetics closure relation of Eq. (\ref{Henry}): 

\begin{equation} \label{box2}
\frac{d \Pi_{\pm}(t)}{dt}  = k_{a} c(\pm L,t) - k_{d} \Pi_{\pm}(t).
\end{equation}

In Appendix E we give both $\tilde{\Pi}(s)$ and the Laplace transform of the density profile. Here, we highlight results for the symmetric initial condition  $x_0=0$, where $\tilde{\Pi}(s)$ simplifies to

\begin{align}  \label{box4}
\tilde{\Pi}(s) =
&\frac{k_{a}}{(k_{a} s) \text{cosh}(L \sqrt{\frac{s}{D}}) + \sqrt{s D} (k_{d} + s) \text{sinh}(L \sqrt{\frac{s}{D}})  }. 
\end{align}

\begin{figure}[t]
\begin{centering}
\includegraphics[width=1\linewidth]{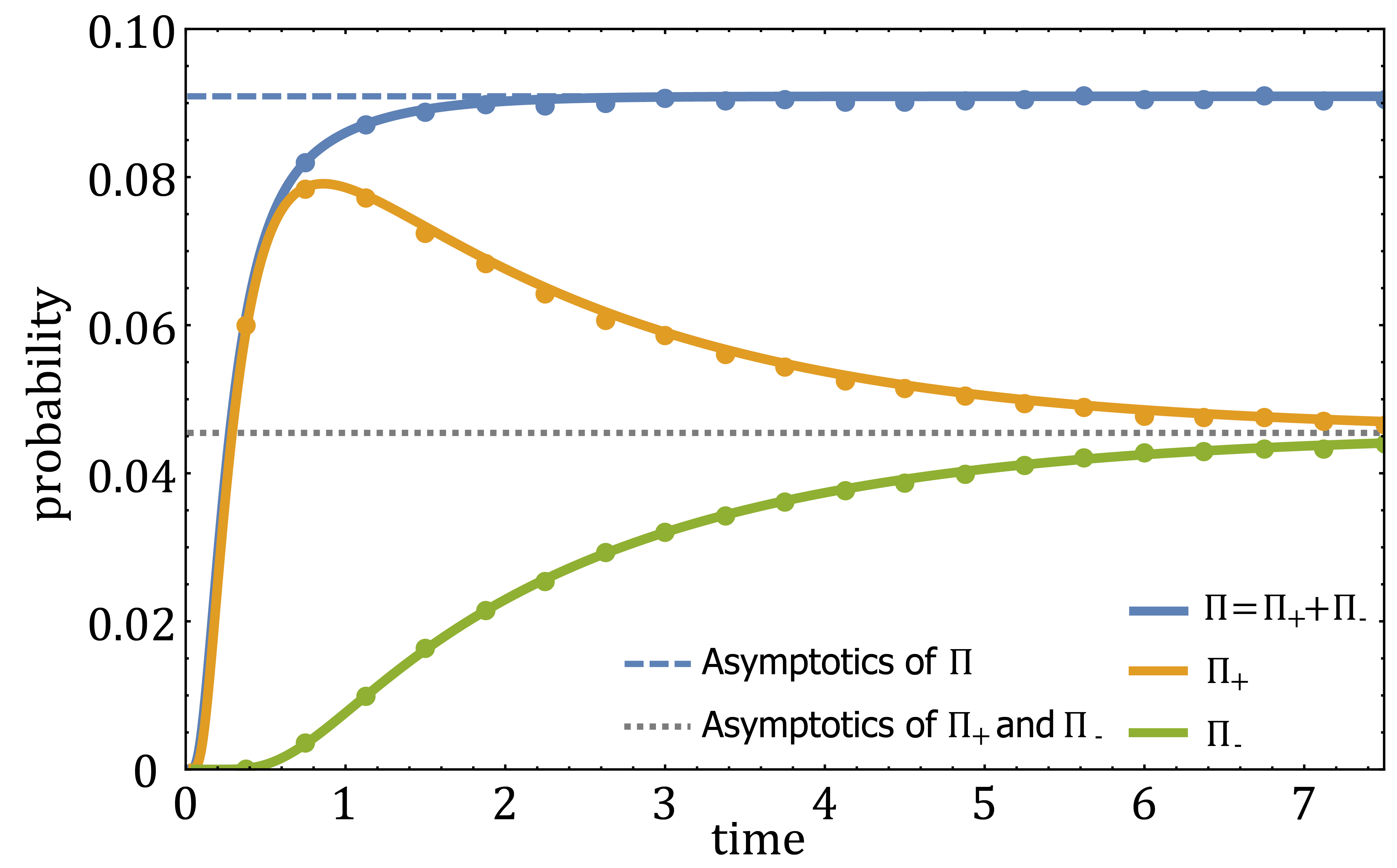}
\caption{An adsorbate which starts at $x_0$ diffuses in an interval of length $2L$ between two identical adsorbing surfaces with linear adsorption kinetics. The solid lines give the probability $\Pi_{+}(t)$ to be on the right surface, the probability $\Pi_{-}(t)$ to be on the left surface, and their sum $\Pi(t)$. These lines are plotted by inverting Eq. (\ref{F:1}) numerically \cite{abate2004multi}. Full circles come from simulations, and the dashed lines are the long time asymptotics of Eqs. (\ref{box5}) and (\ref{box6}). Here, we have used the following parameters: $L=2$, $x_0=1$, $D=1$, $k_a=2$, $k_d=10$.}
\end{centering}
\end{figure}

Consider the long time asymptotics, corresponding to the limit $s \to 0$. By expanding Eq. (\ref{box4}) around $s=0$ and inverse Laplace transforming the first term of the expansion we obtain 

\begin{equation}  \label{box5}
\Pi(t) \simeq \frac{K}{K+L},
\end{equation}

\noindent where $K = k_{a}/k_{d}$. Since the surfaces are taken to be identical and the result in Eq. (\ref{box5}) is for the long-time asymptotics, it is actually independent of the initial condition and is valid for any $x_0$. Also note that in the long time limit the probability to be on either surface is the same 

\begin{equation}  \label{box6}
\Pi_{\pm}(t) \simeq \frac{1}{2}\frac{K}{K+L}.
\end{equation}

\noindent In Fig. 3 we demonstrate these results and compare them to Monte-Carlo simulations.

\subsubsection{Initial uniform concentration}

As previously discussed, all we need to do in order to obtain a solution for a macroscopic adsorption problem is to integrate over the initial position $x_0$ of the single-particle solution with proper weights. In this case, we integrate the solution in Eq. (\ref{F:1}) between $-L$ and $L$ with a constant weighting of $c_b$ (uniform concentration) to obtain

\begin{align}  \label{boxmacro1}
\tilde{\Gamma}(s) =
&\frac{2 k_{a}\operatorname{sinh}( L \sqrt{\frac{s}{D}})}{k_{a}\frac{s^{3 / 2}}{\sqrt{D}} \operatorname{cosh}( L \sqrt{\frac{s}{D}})+\left(s^{2}+k_{d} s \right)\operatorname{sinh}( L \sqrt{\frac{s}{D}})} c_{b}. 
\end{align}

\noindent In Appendix F we also solve this model from scratch. Indeed, the result remains the same.

The long time asymptotics of Eq. (\ref{boxmacro1}), corresponding to the limit $s \to 0$, is given by

\begin{equation}  \label{boxmacro2}
\Gamma(t) \simeq  \frac{K}{K+L} 2 L c_{b},
\end{equation}

\noindent which is the law of Eq. (\ref{box5}) multiplied by $2 L c_b $, namely, the number of particles in the interval divided by the surface area $\mathcal{A}$. The result of Eq. (\ref{boxmacro2}) is an equilibrium result. It suggests that a simple surface concentration measurement at equilibrium is sufficient to determine $K$. Equation (\ref{boxmacro1}) is inverted numerically in Fig. 4, where it is shown to be in agreement with Monte Carlo simulations and the asymptotics of Eq. (\ref{boxmacro2}).

We can also invert Eq. (\ref{boxmacro1}) analytically. The poles are determined by the zeros of the denominator. Setting $\beta=i L\sqrt{\frac{s}{D}}$, so that $s=-\frac{\beta^2 D}{ L^2}$, we find the following equation for $\beta$

\begin{equation}  \label{boxmacr_poles}
\frac{\beta}{\tan(\beta)} = \frac{ D \beta^2 - L^2 k_d  }{ L k_a}. 
\end{equation}

There is an infinite set of solutions of this equation that we denote as $\beta_n$. The first solution $\beta_0=0$ corresponds to the constant term in Eq. (\ref{boxmacro2}), while the others lie inside $((n-1)\pi,n \pi)$ and can be calculated numerically. In turn, the poles are $s_n=-\frac{\beta_n^2 D}{ L^2}$. To find the residues at the poles we first need to calculate the derivative of the denominator in Eq. (\ref{boxmacro1}) with respect to s:

\begin{align}
&\phi'(\beta)=\frac{\partial \phi(\beta)}{\partial s} \frac{\partial s}{\partial \beta} = \\
 &\frac{-i}{2L^2} \times  \Big\{\beta \left(3 L k_a+ L^2 k_d-\beta ^2 D\right) \cos (\beta ) \nonumber\\
&  +  \left[2 L^2 k_d - \beta ^2 \left(L k_a+4 D\right) \right] \sin (\beta )\Big\}   \nonumber.
\end{align}

Applying the residue theorem we obtain 

\begin{align} \nonumber
\Gamma(t) & = \frac{1}{2\pi i}\int\limits_\gamma e^{st} \tilde{\Gamma}(s) ds = \sum\limits_n e^{s_nt} \textrm{Res}_{s_n} \{\tilde{\Gamma}(s)\} \\ 
& = \frac{K}{K+L} 2 L c_{b} +
c_b\sum\limits_{n=1}^\infty e^{-\beta_n^2 Dt/ L^2} \frac{-i 2 k_{a}\operatorname{sin}(\beta_n)}{\phi'(\beta_n)} . \label{uniforminversion}
\end{align}

Equation (\ref{uniforminversion}) can be evaluated, and the obtained solution fits the curve in Fig. 4. The $\tilde{\Gamma}(s)$ probabilities for the rest of the confined cases can be inverted using the same technique \cite{grebenkov2013geometrical}.

While Eq. (\ref{boxmacro2})
determines the asymptotic behaviour, the eigenvalue $s_1$ characterize the long time approach to this equilibrium. Expanding the fraction on the left hand side of Eq. (\ref{boxmacr_poles}) and retaining terms up to first order in $\beta$, when $\beta \to 0$, we obtain $\beta/\tan(\beta) \simeq 1 -\beta^2/3+O(\beta^4).$ Substituting the above expansion and $s=-\frac{\beta^2 D}{ L^2}$ into Eq. (\ref{boxmacr_poles}), and solving for $s$ we obtain

\begin{equation}\label{beta1approx}
s_1 \simeq -  \frac{k_a/+Lk_d}{L+L^2k_a/3D}.   
\end{equation}

\noindent This approximation is valid whenever $\beta_1$ is small, i.e., when the adsorption and desorption rates are small compared to the time it takes to diffuse the channel.

\begin{figure}[t]
\begin{centering}
\includegraphics[width=1\linewidth]{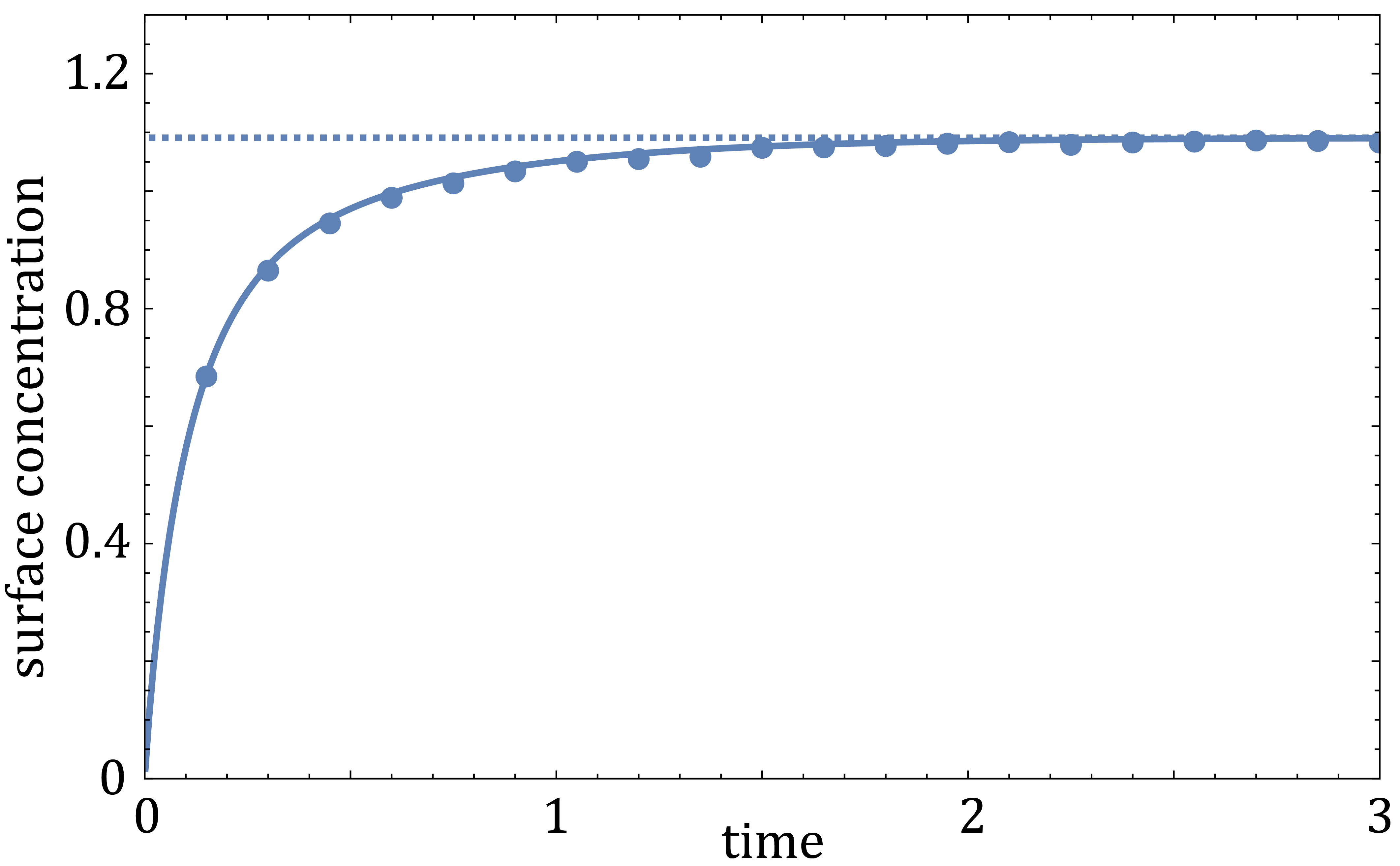}
\caption{Surface concentration vs. time, for a dilute solution of adsorbates that diffuse in an interval of length $2L$ between two identical adsorbing surfaces with linear adsorption kinetics. The solid line gives $\Gamma(t)$, which is plotted by inverting Eq. (\ref{boxmacro1}) numerically \cite{abate2004multi}. The full circles come from simulations, and the dashed line is the long time asymptotics of Eq. (\ref{boxmacro2}). Here, we have used the following parameters: $L=2$, $c_b=3$, $D=1$, $k_a=2$, $k_d=10$. }
\end{centering}
\end{figure}

\section{III. Adsorption kinetics in higher dimensions and general geometries}

\subsection{A. Generalization of the Ward-Tordai relation}

The Ward-Tordai relation is based on a 1-dimensional model, with a specific geometry and initial condition (Fig. 1a). Now that we understand the physics behind it (recapitulated in Eq. (\ref{WardTordaiInTermsOfGreens})), we can easily generalize this relation to higher dimensions, arbitrary geometries and different initial conditions. 

Let $\mathcal{D}$ be a $d$-dimensional domain in which an adsorbate is bounded by a smooth closed surface $S$ (closed in the broad sense, including infinities). Let $p(\Vec{r},t)$ be the probability density of finding a diffusing adsorbate at $\Vec{r}$ at time $t$, given an initial profile $p_{0}(\Vec{r})$. The subsurface probability density is an unspecified time-dependent boundary condition. Let $g(\Vec{r},t \mid \Vec{r}',\tau)$ be the corresponding Green's Function, i.e., the solution to the corresponding absorbing
boundary problem. We denote $\int_{\mathcal{D}}d\Vec{r}$ and $\oint_{S}d\Vec{r}$ as the $d$-dimensional domain integral and the $(d-1)$-dimensional surface integral respectively. We further denote by $ \Vec{n}_{i}$ the inward-drawn normal to the surface, i.e. pointing towards the surface. 

To get the overall probability to be on the surface we integrate over all $\Vec{r}_s\in S$ and obtain $\Pi(t):=\oint_{S}\Pi(\Vec{r}_{s},t) d\Vec{r}_{s}$. 
We can write

\begin{equation}  \label{GeneralizedWardTordai}
\begin{array}{ll} 
\Pi(t) = \underbrace{1 - \int_{\mathcal{D}} \left[\int_{\mathcal{D}}g(\Vec{r},t \mid \Vec{r}',0)  p_{0}(\Vec{r}') d \Vec{r}' \right] d \Vec{r} }_{\textrm{First term}}
\\
\\
-\underbrace{D \int_{\mathcal{D}}\int_{0}^{t}\left[\oint_{S}p(\Vec{r}',\tau) \frac{\partial g(\Vec{r},t \mid \Vec{r}',\tau)}{\partial \Vec{n}'_{i}} d \Vec{r}'\right] d \tau d\Vec{r}}_{\textrm{Second term}},
\end{array}
\end{equation} 

\noindent where we again used $Q(t) = 1 - F(t)$, such that the first term is written in terms of the Green's function itself, rather than in terms of its derivative.

The first term gives the overall probability to find the adsorbate on the surface in case that the surface is completely adsorbing ($k_{a} \to \infty$) and there is no desorption taking place ($k_{d} \to 0$). This probability is then equivalent to the probability that a particle is absorbed by a perfectly absorbing surface, which can be easily written in terms of the Green's function (we calculate the cumulative of the higher dimensional analogue of the first-passage distribution as expressed in Eq. (\ref{firstpassage})). The second term is a corrective term that accounts for the adsorption dynamics, at every time point $\tau \in [0,t)$ prior to the observation time. We do so by setting a subsurface probability density $p(\Vec{r},\tau)$ for every $\Vec{r} \in S$ and letting it evolve an additional $t-\tau$ time units according to a doublet propagator up to the observation time (note that time-reversal symmetry arguments can also explain this quantity \cite{grebenkov2019spectral,grebenkov2020paradigm}). We then integrate over the surface and over $\tau$ to get the contributions from the entire surface at all times. Finally, we integrate over the entire domain $\mathcal{D}$ to get the overall probability that the adsorbate started at the subsurface at time $\tau$ and ended up in the bulk at time $t$. This amount is the second term that needs to be subtracted from the first.

Finally, note that all we need in order to express $\Pi(t)$ are the initial probability density, the subsurface probability density and the Green's function. In Appendix G we give a bottom-up derivation of Eq. (\ref{GeneralizedWardTordai}) for the case of 3-dimensional domain that is based on the work of Carslaw and Jaeger \cite{carslaw1959conduction}. The derivation can be easily adapted to lower dimensions.

Adsorption problems in higher dimensions differ significantly from the 1-dimensional theory presented above in one crucial point -- the surface is no longer comprised of a single point, and so we have to treat each surface point independently. Unfortunately, it seems that a general Ward-Tordai like relation for the probability density to be on the surface at a specific point $\Vec{r}_s$ can only be written in a circular manner which involves all the other points on the surface. Note, however, that for radially symmetric problems, the probability to be at any surface point is the same, and so $\Pi(\Vec{r}_{s},t)=\Pi(t)/\oint_{S} d\Vec{r}_{s}$. An example for such a case is that of diffusion inside an adsorbing spherical shell, which is considered next.


\subsection{B. Diffusion inside an adsorbing spherical shell}
As an example for a higher dimensional problem, consider the model illustrated in Fig. 5. An adsorbate diffuses inside a spherical shell, where the radii of the inner and outer spheres are $R_-$ and $R_+$ respectively and the initial distance of the adsorbate from the origin is $r_0$. Both spheres are taken to be homogeneously adsorbing, with $k_{a}$ and $k_{d}$ standing for the adsorption and desorption rate constants. The probability to find the particle somewhere on the inner/outer sphere of radius $R_{\pm}$ at time $t$ is denoted by $\Pi_{\pm}(t)$. By taking the limit $R_- \to 0$ we get the model of an adsorbate diffusing inside an homogeneously adsorbing sphere. Similarly, by taking the limit $R_+ \to \infty$ we get the model of an adsorbate diffusing outside a homogeneously adsorbing sphere (this limit corresponds to the Smoluchowski-type reaction solved by Kim and Shin \cite{kim1999exact}).

\begin{figure}[t]
\begin{centering}
\includegraphics[width=0.8\linewidth]{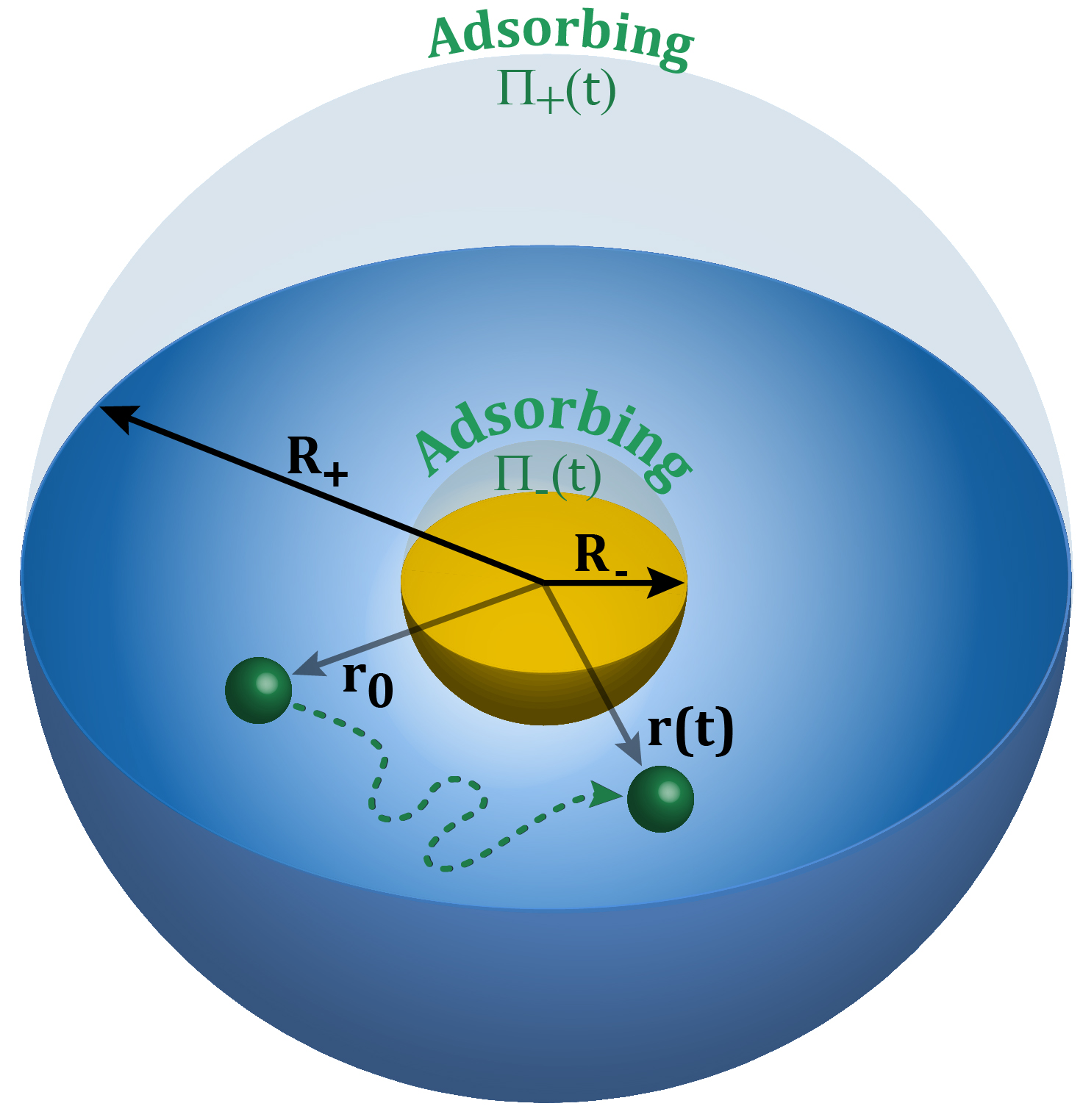}
\caption{Illustration of an adsorbate diffusing inside a spherical shell, where the radii of the inner and outer spheres are $R_-$ and $R_+$ respectively and the initial distance of the adsorbate from the origin is $r_0$. Both spheres are taken to be homogeneously adsorbing, with $k_{a}$ and $k_{d}$ as the adsorption and desorption rate constants. The probability to find the particle somewhere on the sphere of radius $R_{\pm}$ at time $t$ is denoted by $\Pi_{\pm}(t)$. The probability to find the adsorbate a distance $r$ from the origin at time $t$ is given by $p(r,t)$. By taking the limit $R_- \to 0$ we get the model of an adsorbate diffusing inside an homogeneously adsorbing sphere. Similarly, by taking the limit $R_+ \to \infty$ we get the model of an adsorbate diffusing outside a homogeneously adsorbing sphere.}
\end{centering}
\end{figure}

Starting with the equation for a freely diffusing particle in 3-dimensions, we replace the 1-dimensional Laplacian in Eq. (\ref{Micro1}) with a 3-dimensional Laplacian. Furthermore, we utilize the rotational symmetry of the model and consider the Laplacian in spherical coordinates, where the propagator is independent on the angular part of the Laplacian. Denoting the distance from the origin by $r \equiv \abs{\vec{r}}$, we are left only with the radial part of the Laplacian $\frac{1}{r^{2}}\frac{\partial}{\partial r} \left( r^{2} \frac{\partial}{\partial r}\right)$. The diffusion equation can thus be written as

\begin{equation} \label{radial_diffusion}
    \frac{\partial  p(\vec{r}, t)}{\partial t}  =  \frac{2D}{r} \frac{\partial p(\vec{r}, t )}{\partial r}
   +D \frac{\partial^{2} p(\vec{r}, t)}{\partial r^{2}}. 
\end{equation}

Equation (\ref{radial_diffusion}) is written for the propagator. In what follows, we will be interested in the probability density to be at a distance $r$ from the origin: $p(r, t) =4 \pi r^{2} p(\vec{r}, t) $, where $4 \pi r^2$ is the surface area of a 3-dimensional sphere of radius $r$ (essentially, we are studying the 3-dimensional Bessel process \cite{bray2000random,martin2011first,ryabov2015brownian,ray2020diffusion} with adsorption). By plugging this relation into  Eq. (\ref{radial_diffusion}) we obtain the Fokker-Planck equation for $p(r, t)$:

\begin{equation} \label{radial1}
 \frac{\partial  p(r, t )}{\partial t}  =   \frac{\partial}{\partial r}\Big[\left(-\frac{2 D}{  r}\right) p(r, t)\Big]+D \frac{\partial^{2} p(r, t)}{\partial r^{2}}, 
\end{equation}

\noindent with the following initial conditions

\begin{equation}  \label{radial2}
\begin{array}{ll}
     p(r,0) =  \delta(r-r_0),                 \text{\hspace{5.75ex}(a)}   \\
     \Pi(0)= \Pi_{\pm}(0)=0.        \text{\hspace{5.5ex}(b)}
\end{array}
\end{equation}

The mass-balance boundary conditions are also slightly modified in spherical coordinates

\begin{equation}  \label{radial3}
  \frac{d \Pi_{\pm}(t)}{dt} = J(R_{\pm},t) = \mp \left[D\frac{\partial p(r, t)}{\partial r} - \frac{2 D}{r}p(r, t)\right]_{r=R_{\pm}},
\end{equation}

\noindent where $J(R_{\pm},t)$ are the fluxes into the inner and outer adsorbing spheres. Finally, as before, we assume linear adsorption kinetics on the surfaces so the closure relations are the same as in Eq. (\ref{box2}) just with $R_{\pm}$ instead of $\pm L$. 

\begin{figure}[t!]
\begin{centering}
\includegraphics[width=1\linewidth]{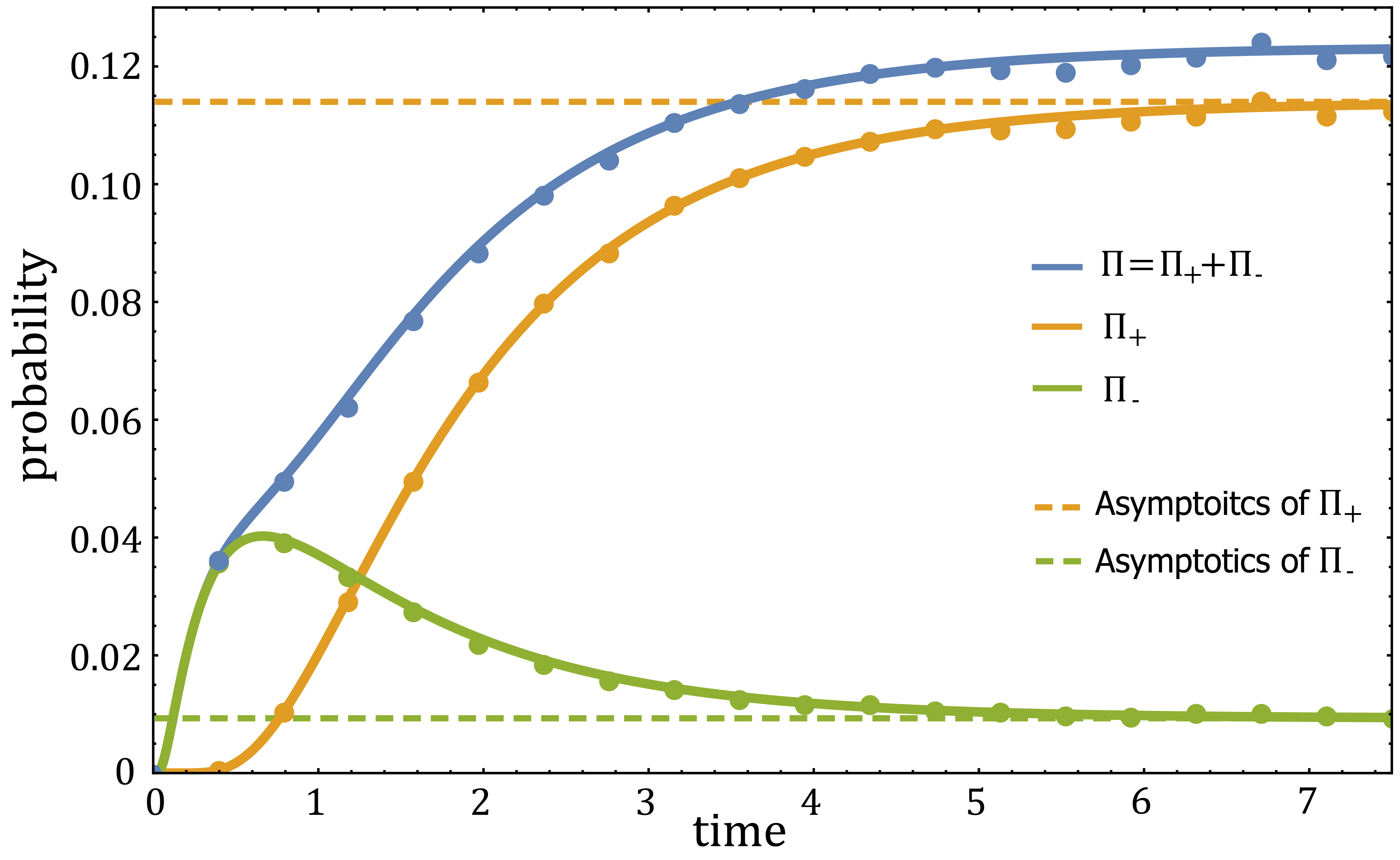}
\caption{The probabilities of finding the particle on the inner, $\Pi_{-}(t)$, or outer, $\Pi_{+}(t)$, surfaces and their sum $\Pi(t)$  vs. time. The solid lines were plotted using a numerical inverse Laplace transformation of Eq. (\ref{H:1}) \cite{abate2004multi}, for an adsorbate that diffuses in a spherical shell with inner radius $R_{-} = 2$ and outer radius $R_+=7$. Here, we assume that the adsorbate starts at a distance $r_0 = 3$ from the origin, and that two adsorbing surfaces are identical with respect to the adsorption kinetics, which is linear with $k_a = 0.8$ and $k_d = 2.7$. The diffusion coefficient was taken to be $D=2.1$. Full circles come from simulations, and the dashed lines are the long time asymptotics coming from Eq. (\ref{radial6}). }
\end{centering}
\end{figure}

To solve, we Laplace transform Eq. (\ref{radial1}) and obtain 

\begin{align}\label{radial4}
   \frac{d^{2} \tilde{p}(r, s)}{d r^{2}} - & \frac{d}{d r}\Big[\left(\frac{2}{ r}\right) \tilde{p}(r, s)\Big]\\
   &- \frac{s}{D}\tilde{p}(r, s) + \frac{\delta(r-r_0)}{D}=0 ,  \nonumber
\end{align}

\noindent which is a second order ordinary differential equation.  The general solution for $\tilde{p}(x, s)$ is \cite{redner2001guide,ray2020diffusion}

\begin{align} \label{radial5}
\tilde{p}(r, s) = & r^{\frac{3}{2}} \times \\
&\begin{cases}A_{1} I_{ -\frac{1}{2}}\left(\sqrt{\frac{s}{D}} r\right)+B_{1} K_{ -\frac{1}{2}}\left(\sqrt{\frac{s}{D}} r\right) ,& r > r_{0} \\ A_{2} I_{ -\frac{1}{2}}\left(\sqrt{\frac{s}{D}} r\right)+B_{2} K_{-\frac{1}{2}}\left(\sqrt{\frac{s}{D}} r\right) ,& r<r_{0}\end{cases}, \nonumber
\end{align}

\noindent where $I_{-\frac{1}{2}}(\cdot)$ and $K_{-\frac{1}{2}}(\cdot)$ are the modified Bessel functions of the first and second kinds of order $-1/2$. To determine the coefficients $A_1(s), A_2(s), B_1(s), B_2(s)$ we impose the Laplace transform of the above boundary conditions and the Laplace transform of the matching (continuity) condition at $r_0$. Yet, even in Laplace space, the coefficients in Eq. (\ref{radial5}) and $\tilde{\Pi}(s)_{\pm}$ are too bulky to be presented here and they are thus relegated to Appendix H. 

Since the particle is restricted between two spheres, a steady state will eventually be attained. Taking $s \to 0$, and inverting the result, we obtain the long time asymptotics of the probability to be adsorbed to the surfaces

\begin{equation}  \label{radial6}
\Pi_{\pm}(t) \simeq \frac{3 k_a R_{\pm}^2}{3 k_a (R_{+}^2+R_{-}^2)+k_d (R_{+}^3 - R_{-}^3)}.
\end{equation}

\noindent In Fig. 6 we demonstrate these results and compare them to Monte-Carlo simulations.

Of course, as in the previous examples, taking the limits $k_a \to 0$ or $k_d \to \infty$ results in $\Pi_{\pm}(t) = 0$. In the limit of $k_a \to \infty$ we obtain a compact formula in Laplace space

\begin{align}  \label{radial7}
\tilde{\Pi}_{\pm}(s) & = \\
&\mp \frac{  R_{\pm} \text{csch}\left[-\sqrt{\frac{s}{D}} (R_+ - R_-)\right]  \text{sinh}\left[-\sqrt{\frac{s}{D}} (R_+ - r_0)\right] }{s r_0}. \nonumber
\end{align}

\noindent The formulas in the limit $k_d \to 0$ are again bulky and are not given here.



As mentioned above, by taking the limits $R_- \to 0$ in Eq. (\ref{H:1}) the problem reduces from diffusion in an adsorbing spherical shell to that of diffusion inside an adsorbing sphere of radius $R_+$. Similarly, by taking the limits $R_+ \to \infty$ in Eq. (\ref{H:2}) the problem reduces to that of diffusion outside an adsorbing sphere of radius $R_-$. We will now present results for both these special cases.

\begin{widetext}
\subsubsection{Diffusion inside an adsorbing sphere}

By taking the radius of the inner sphere to zero, $R_- \to 0$ in Eq. (\ref{H:1}), we obtain

\begin{equation}  \label{insidesphere1}
\tilde{\Pi}_{+}(s) = \frac{ k_a R_+^2 }{r_0}\frac{ \sinh \left(r_0
   \sqrt{\frac{s}{D}}\right)}{ R_+ \sqrt{D
   s} \left(k_d+s\right) \cosh \left(R_+
   \sqrt{\frac{s}{D}}\right)-\sinh \left(R_+
   \sqrt{\frac{s}{D}}\right) \left[D
   \left(k_d+s\right)-s k_a R_+ \right]}.
\end{equation}

By taking this limit in Eq. (\ref{radial6}) we obtain the probability to be found adsorbed to the sphere at long times

\begin{equation}  \label{insidesphere2}
\Pi_+(t) \simeq \frac{3 K}{3 K + R_+}.
\end{equation}

\subsubsection{Diffusion outside an adsorbing sphere}

By taking the radius of the outer sphere to infinity, $R_+ \to \infty$ in Eq. (\ref{H:2}), we obtain

\begin{equation}  \label{outsidesphere1}
\tilde{\Pi}_{-}(s) = \frac{k_a R_-^2}{r_0} \frac{e^{-\sqrt{\frac{s}{D}} \left(r_0-R_-\right)}
   \left(s k_a+k_d \sqrt{D s}+\sqrt{D
   s^3}\right)}{\sqrt{s} \left[\sqrt{s} k_a+\sqrt{D}
   \left(k_d+s\right)\right] \left[R_- \left(s
   k_a+\sqrt{D s} \left(k_d+s\right)\right)+D
   \left(k_d+s\right)\right]}.
\end{equation}
\end{widetext}

This case is similar to the example of diffusion on the semi-infinite line, in the sense that steady-state is never attained. There is, however, a big difference coming from the fact that a 3-dimensional random walk is not  recurrent \cite{redner2001guide}. Thus, over an infinite measurement horizon, the particle will only spend a finite amount of time adsorbed (while for the 1-dimensional case this time diverges). Let us denote the ensemble average of this time duration by

\begin{equation}  \label{outsidesphere2}
\Lambda_{\infty} := \lim_{t \to \infty}\Lambda(t) = \int_0^\infty \Pi_{-}(t') dt' = \tilde{\Pi}_{-}(0)  ,
\end{equation}

\noindent where $\tilde{\Pi}_{-}(0)$ is the Laplace transform of $\Pi_{-}(t)$ evaluated at $s=0$. Thus, taking the limit $s \to 0$ in Eq. (\ref{outsidesphere1}), we obtain

\begin{equation}  \label{outsidesphere3}
\Lambda_{\infty} =  \frac{K R_-^2}{D r_0}.
\end{equation}
 
It can be seen that whenever $K \to 0$ (the limit of a reflecting surface), the time spent on the target naturally decreases. Similarly, as the adsorbate starts further from the surface (larger $r_0$) or for a smaller adsorbing sphere (smaller $R_-$), $\Lambda_{\infty}$ decreases. Finally, $\Lambda_{\infty}$ decreases as $D$ increases. This is because the particle spends less time in the subsurface for larger diffusion coefficients, and consequently the probability to be adsorbed during this time becomes smaller.

Finally, in Appendix I, we obtain the following long-time asymptotics

\begin{equation}  \label{outsidesphere4}
\Pi_{-}(t) \simeq \frac{K R_-^2}{\sqrt{4 \pi D^3}}t^{-3/2},
\end{equation}

\noindent which should be compared with Eq. (\ref{surface2}) of the analogous 1-dimensional problem. 

The problem of diffusion outside an adsorbing sphere is mappable onto  Smulochowski-type reversible reactions. Results corresponding to Eqs. (\ref{outsidesphere1})-(\ref{outsidesphere4}) can be found in the work of Kim and Shin \cite{kim1999exact}. Notably, they have further provided an analytic inversion of Eq. (\ref{outsidesphere1}). 
\section{IV. Discussion}

\subsection{A. Experimental considerations}
A major challenge in macroscopic adsorption kinetics studies is the proper estimation of the surface concentration. For liquid-air interfaces, as well as for liquid-liquid interfaces, this is usually done indirectly via measurement of the surface tension \cite{lin1990diffusion}. For an ideal dilute solution one then uses the Gibbs adsorption equation to relate the surface tension $\gamma$ to the surface excess $\Gamma_{ex} = - \frac{1}{RT} \left( \frac{\partial \gamma}{\partial \text{ln} c_b}\right)_{T}$, where $R$ is the gas constant, $T$ is the temperature, and $c_{\mathrm{b}}$ is the concentration in the bulk  \cite{chattoraj1984miscellaneous}. However, for non-ideal conditions or for out-of-equilibrium measurements, the exact relation between these two quantities is unclear \cite{chang1995adsorption,liu2000diffusion}. Furthermore, the surface phase might be of non-negligible width and display inhomogeneous adsorbate concentration. Such complications make it harder to infer the equilibrium surface concentration $\Gamma_{eq}$ from the surface excess $\Gamma_{ex}$ \cite{chattoraj1984miscellaneous,wang2014dependence}. While single-particle experiments are significantly more difficult to conduct, they are usually done in a direct manner (e.g., via electron or light microscopy or via atomic force microscopy), and it should be relatively straight forward to interpret whether the molecule is on the surface or not. 

However, in single-particle experiments other problems may arise. For example, the need for tracking of a single particle throughout the sample, and not merely on the surface, in order to make sure that the studied particle is not replaced by another while leaving and returning to the field of view. Perhaps performing multiple experiments can be the remedy here. For instance, one can evaluate the desorption rate $k_d$ separately in a well controlled single-molecule experiment, and later deduce the adsorption rate $k_{a}$ from a macroscopic equilibrium measurement of a dilute solution.

A clear advantage of single-particle experiments is the very plausible circumvention of Langmurian dynamics, as such experiments will be naturally conducted in extreme dilution, thus making surface saturation effects negligible. While these effects are the main obstacle for analytical analysis of macroscopic adsorption kinetics, it seems that for single-particle kinetics they can be safely ignored and Henry (linear) dynamics are sufficient.

Furthermore, if one still wishes to challenge the linear kinetics assumption, single-particle adsorption experiments open a new route for doing so. Implicit in the linear dynamics assumption hides the Markovian assumption, i.e., that both adsorption and desorption times are taken from an exponential distribution with mean times of $k_{a}^{-1}$ and $k_{d}^{-1}$. Perhaps in reality these times are distributed according to other, i.e. non-exponential, distributions. To this end, the work of Agmon and Weiss on dissociation-recombination reactions \cite{agmon1989theory} stands out to be an excellent reference.

\subsection{B. Discussion and Outlook}

In this work we have presented a complete and self-contained account of adsorption kinetics, from macro- to microscopic dynamics. Notably, we have shown how the former can be generalized from the latter one. One noteworthy important result of our work is the generalization of the Ward-Tordai relation to any dimension, geometry and initial conditions. This generalization was built side by side with a detailed interpretation of the Ward-Tordai relation, which was aimed to build strong intuition for the physics of the problem. From a technical point of view, we have shown that the Ward-Tordai relation can be written in terms of the Green's function, which is the fundamental solution to the problem with an absorbing, rather than adsorbing, boundary. This is a crucial aspect to this work.

We further showcased the results obtained by analytically solving fundamental adsorption kinetics models that were not studied before -- 1-dimensional diffusion inside a interval with adsorbing boundaries, and 3-dimensional diffusion inside an adsorbing sphere/spherical shell. We hope that the general discussion presented herein will inspire others to explore additional models which are governed by more complicated mass-transport processes, encompassing different surface morphologies, equipped with non-Markovian adsorption and desorption dynamics, and so on.

The theory that was presented herein decouples between the mass-transport to the surface and the adsorption kinetics actually taking place at the surface. We argue that this separation is not only a consequence of the mathematical approach we have chosen, but also is conceptually important. Indeed, when experimentally reporting on the adsorption and desorption rates $k_a$ and $k_d$, the measurement usually embodies the overall rates, with contributions coming from both processes. We believe that, if possible, these rates should encapture only the adsorption once at the subsurface, and the rate of desorption back to the subsurface \footnote{When defined this manner, these rates are sometimes referred to as $k_{\textrm{on}}$ and $k_{\textrm{off}}$, but $k_a$ and $k_d$ seem to be preferred in some disciplines, so that adopting a new terminology may end up causing more confusion. One should simply make the definition of the rate constants very clear.}. 

Adsorption-desorption dynamics are similar in nature to reversible reaction (recombination-dissociation) dynamics, for which macro- and microscopic theories were developed and thoroughly investigated
\cite{agmon1984diffusion,agmon1988geminate,agmon1990theory,kim1999exact,agmon1993competitive,gopich1999excited,prustel2012exact}. In fact, the Henry (linear) closure relation of absorption dynamics can be identified as the ``back reaction" boundary condition of recombination dynamics which was introduced by Agmon \cite{agmon1984diffusion}. 

Agmon's solution for the problem of 1-dimensional reversible reaction on the semi-infinite interval can be mapped to the solution of the semi-infinite 1-dimensional adsorption problem that we have solved and analyzed in Sec. II.C \cite{agmon1984diffusion,agmon1988geminate}. Similarly, diffusion $\textit{outside}$ of an adsorbing sphere is mappable to the reversible reaction model considered by Shin and Kim \cite{kim1999exact}. An additional analytically solvable model was first suggested by Prüstel and Tachiya, and further investigated by Grebenkov: a reversible reaction where the particle is restricted to diffuse on a 3-dimensional sphere towards an reversibly-reacting/adsorbing circle \cite{prustel2013reversible,grebenkov2019reversible}.

Adsorption-desorption and recombination-dissociation dynamics are, however, not identical. The theory of recombination-dissociation reactions can be viewed as a generalization of  Smoluchowski's spherically symmetric diffusion-influenced reaction model \cite{collins1949diffusion}. When studying an isolated pair of particles according to this model, one sets the origin of the coordinates system at the center of particle A, assumed spherical, such that the coordinates system moves with the particle. Particle B, assumed spherical as well, then appears to be diffusing with a diffusing coefficient $D=D_A+D_B$. A recombination reaction can take place when the two particles are at contact, i.e., when the reaction coordinate is at a distance $R_A+R_B$ from the origin (where $R_i$ is the radius of particle $i$), according to the ``back reaction" boundary condition. If a reaction does not take place, the particles are reflected. Thus, by using this model one considers only closed spherical surfaces, where the particle diffuses \textit{outside} of the adsorbing sphere. For this reason, scenarios where adsorption occurs on the boundaries of an interval, box, cylinder, or spherical shell were not studied in the context of recombination-dissociation dynamics.

In contrast, the classical model of adsorption kinetics is that of a particle that adsorbs to a flat surface, with no spherical symmetry (apart from the redundant 1-dimensional case), and usually the surface is envisioned to be much larger than the particle. That said, the generalized Ward-Tordai relation given in Eq. (\ref{GeneralizedWardTordai}) does not impose any morphological restriction on the surface, other than being closed (in the broad sense, including infinities). For example, in Sec. III we investigated the motion of a particle that diffuses \textit{inside} a spherically symmetric adsorbing surface, which can perhaps be of use when considering adsorption kinetics within living or artificial cells.

In works on recombination-dissociation reactions an additional component is routinely added to the model -- the particles, which are assumed to be excited, can later on decay to their lower energy state. Excited particles thus have a lifetime and the decayed/quenched particles are not counted/observed. In the same spirit, one can extend the adsorption-desorption dynamics discussed herein to include non-conservative reactions, e.g., by adding a rate of decomposition for molecules that are adsorbed to the surface (as was done for macroscopic reactions catalyzed by solid surfaces \cite{house2007principles}). Another technique that can be borrowed is the very recently applied renewal approach to reversible binding reactions \cite{reva2021first}, which is in fact general in its derivation. By employing such a technique one forgoes the calculation of the full propagator in favour of the surface occupancy probability. However, as calculation of the propagator can be extremely difficult, a circumventing route is sometimes favourable.

Lastly, we wish to draw attention to a recent line of work which attempts to model reactions that are triggered upon a threshold crossing event, using multiple particles with reversible-binding kinetics \cite{grebenkov2017first,lawley2019first,grebenkov2021reversible}. In the first work of this series \cite{grebenkov2017first}, Grebenkov formulated the problem of ``impatient particles'': given $N$ independent particles diffusing in confinement in the presence of an adsorbing target (e.g., a surface, a sensor, and so on), when will $K$ particles be bound to the target for the first time? Though the problem was posed in full generality, Grebenkov introduced a way to calculate the mean reaction time for the case of $N=K=2$ \cite{grebenkov2017first}. Subsequently, Lawely and Madrid proposed an  approximation (very accurate for small targets with low reactivity) through which one can model the dynamics of the number of particles bound to the target as a Markovian birth-death process \cite{lawley2019first}. Very recently, Grebenkov and Kumar showed how a renewal approach can be employed to get the exact analytical distribution for the case $N=K$ \cite{grebenkov2021reversible} and an approximate solution, complementary to the Lawley-Madrid approach, for $N<K$ \cite{grebenkov2022first}. Though the targets in these examples are usually thought of as some biological sensor (e.g., diffusion towards calcium sensing protein inside a presynaptic bouton), it seems that the same renewal techniques can also be applied to study surface and adsorption phenomena from the type we have considered here. This line of investigation will be further pursued elsewhere.   


\textit{Acknowledgments.}--- The authors wish to thank Haim Diamant, Aanjaneya Kumar, Ella Borberg  and Fernando Patolsky for inspiring discussions. Arnab Pal gratefully acknowledges research support from the Department of Science and Technology, India, SERB Start-up Research Grant Number SRG/2022/000080 and Department of Atomic Energy, India. Shlomi Reuveni acknowledges support from the Israel Science Foundation (grant No. 394/19). This project has received funding from the European Research Council (ERC) under the European Union’s Horizon 2020 research and innovation program (Grant agreement No. 947731). 
\appendix
\renewcommand{\theequation}{A\arabic{equation}}
\renewcommand{\thefigure}{A\arabic{figure}}
\renewcommand{\thesection}{A\Roman{section}} 
\section{Appendix A: Derivation of the Single-Particle Ward-Tordai Relation by Laplace Transformation} \setcounter{equation}{0}
\subsection{Initial position is in the bulk}

We begin by Laplace transforming the conditions in Eqs. (\ref{Micro2}-\ref{Micro4}) of the main text. Equation (\ref{Micro2}.a) can be replaced by two matching conditions, one for the continuity of the Laplace transform of the densities at the initial position of the particle

\begin{equation}  \label{A:1}
     \tilde{p}_{+}(0,s) = \tilde{p}_{-}(0,s),  
\end{equation}

\noindent and one for the Laplace transform of the fluxes 

\begin{equation}  \label{A:2}
     -1=D\Big[\frac{d \tilde{p}_{+}(x,s)}{d x}\big|_{x=0}  - \frac{d \tilde{p}_{-}(x,s)}{d x}\big|_{x=0}\Big], 
\end{equation}

\noindent which is obtained by integrating both sides of the transformed diffusion equation (Eq. \ref{Micro1}) over an infinitesimally small interval surrounding the initial position. Note that the $1$ on the left-hand side comes from the Laplace transform of the delta function initial condition in Eq. (\ref{Micro2}.a). In addition, the Laplace transform of Eq. (\ref{Micro3}) is

\begin{equation}  \label{A:3}
     \tilde{p}_{-}(x \to -\infty,s) = 0,
\end{equation}

\noindent and that of Eq. (\ref{Micro4}) is

\begin{equation}  \label{A:4}
     s \tilde{\Pi}(s) =-D\frac{d \tilde{p}_{+}(x,s)}{d x}\big|_{x=L},
\end{equation}

\noindent where in the Laplace transformation of the time derivative of $\Pi(t)$ we have used the initial condition in Eq. (\ref{Micro2}.b).

After Laplace transforming the diffusion equation (Eq. \ref{Micro1}) we get

\begin{equation} \label{A:5}
   s \tilde{p}(x,s) = D\frac{d^2 \tilde{p}(x,s)}{dx^2}, 
\end{equation}

\noindent which is a second-order, linear, homogeneous differential equation.  It has a general spatial coordinate-dependent solution

\begin{equation}  \label{A:6}
\tilde{p}(x,s)=
\begin{cases}
  \tilde{p}_{-}(x,s) = C_{1}(s) e^{x\sqrt{\frac{s}{D}}} + C_{2}(s) e^{-x\sqrt{\frac{s}{D}}}, \hspace{5ex} x<0    
      \\
  \tilde{p}_{+}(x,s) = C_{3}(s) e^{x\sqrt{\frac{s}{D}}} + C_{4}(s) e^{-x\sqrt{\frac{s}{D}}}. \hspace{5ex} x>0         
\end{cases}
\end{equation}

\noindent Imposing the conditions in Eqs. (\ref{A:1}-\ref{A:4}) on Eq. \ref{A:6}, produces a system of four equations with five unknowns ($\tilde{\Pi}(s)$ and $C_{i}(s)$ for $1 \leq i \leq 4$). However, since we have more unknowns than equations the best we can do is find a relation between the unknowns:

\begin{equation}  \label{A:9}
\begin{cases}
   C_{1}(s) = \frac{1+e^{-2 L \sqrt{\frac{s}{D}}}}{2 \sqrt{D s}}-\sqrt{\frac{s}{D}} e^{-L \sqrt{\frac{s}{D}}} \tilde{\Pi}(s) ,
\\
    C_{2}(s) = 0  ,
\\
    C_{3}(s) = \frac{e^{-2 L \sqrt{\frac{s}{D}}}}{2 \sqrt{D s}}-\sqrt{\frac{s}{D}} e^{-L \sqrt{\frac{s}{D}}} \tilde{\Pi}(s)  ,
\\
    C_{4}(s) = \frac{1}{2\sqrt{D s}}. 
\end{cases}
\end{equation}

Using the fact that $\tilde{p}(L,s)=C_{3}(s) e^{L\sqrt{\frac{s}{D}}} + C_{4}(s) e^{-L\sqrt{\frac{s}{D}}}$, and after some algebra, we get 

\begin{equation} \label{A:7}
  \tilde{\Pi}(s) =  \frac{e^{-L\sqrt{\frac{s}{D}}}}{s}-\sqrt{\frac{D}{s}}\tilde{p}(L,s),
\end{equation}

\noindent which is the Laplace transform of the single-particle analogue of the Ward-Tordai relation.

Note that the propagator in Eq. (\ref{A:6}) can also be expressed in terms of $\tilde{p}(L,s)$ 

\begin{equation}  \label{A:8}
\begin{cases}
   C_{1}(s) = \frac{ 1  - e^{-2 L \sqrt{\frac{s}{D}}}}{2\sqrt{D s}} +  \tilde{p}(L,s) e^{-L \sqrt{\frac{s}{D}}} ,
\\
    C_{2}(s) = 0  ,
\\
    C_{3}(s) = -\frac{e^{-2 L \sqrt{\frac{s}{D}}}}{2 \sqrt{Ds}} + \tilde{p}(L,s) e^{-L \sqrt{\frac{s}{D}}}  ,
\\
    C_{4}(s) = \frac{1}{2\sqrt{D s}}.
\end{cases}
\end{equation}

\subsection{Initial position is on the surface}

In this case, there is no need to divide the propagator using restrictions, and the general solution is given by

\begin{equation}\label{A:10}
  \tilde{p}(x,s) = C_{1}(s) e^{x\sqrt{\frac{s}{D}}} + C_{2}(s) e^{-x\sqrt{\frac{s}{D}}}.
\end{equation}

While the Laplace transform of Eq. (\ref{Micro3}) is unchanged, the Laplace transform of Eq. (\ref{Micro4}) is

\begin{equation}  \label{A:11}
     s \tilde{\Pi}(s) -1 =-D\frac{d \tilde{p}(x,s)}{d x}\big|_{x=L},
\end{equation}

\noindent where in the Laplace transformation of the time derivative of $\Pi(t)$ we have used the initial condition in Eq. (\ref{Micro6}.b). Imposing the conditions in Eqs. (\ref{A:3}) and (\ref{A:11}) on Eq. (\ref{A:10}), produces a system of two equations with three unknowns $(\tilde{\Pi}(s)$, $C_{1}(s)$ and $C_{2}(s))$.

Using the fact that $\tilde{p}(L,s)=C_{1}(s) e^{L\sqrt{\frac{s}{D}}} + C_{2}(s) e^{-L\sqrt{\frac{s}{D}}}$, and after some algebra, we get

\begin{equation} \label{A:12}
  \tilde{\Pi}(s) = \frac{1}{s} -\sqrt{\frac{D}{s}}\tilde{p}(L,s),
\end{equation}
 
 \noindent and 

\begin{equation}  \label{A:13}
\begin{cases}
   C_{1}(s) =  \tilde{p}(L,s) e^{-L \sqrt{\frac{s}{D}}} = \sqrt{\frac{s}{D}} e^{-L \sqrt{\frac{s}{D}}} \left[\frac{1}{s}-\tilde{\Pi}(s)\right],
\\
    C_{2}(s) = 0,
\end{cases}
\end{equation}

\noindent such that the propagator can be written simply as (recall $x<L$)

\begin{equation}  \label{A:14}
  \tilde{p}(x,s) = \sqrt{\frac{s}{D}} e^{(x-L) \sqrt{\frac{s}{D}}} \left[\frac{1}{s}-\tilde{\Pi}(s)\right] .
\end{equation}


\renewcommand{\theequation}{B\arabic{equation}}
\renewcommand{\thefigure}{B\arabic{figure}}
\renewcommand{\thesection}{B\Roman{section}} 

\section{Appendix B: Time integration of the doublet}
\setcounter{equation}{0}

Following p.272 of Carslaw and Jaeger's \textit{Conduction of Heat in Solids} (1959, Oxford University Press), we ask what should be the magnitude $C(t)$ of a continuously applied doublet with the surface in its middle, such that it will be equivalent to a source of magnitude $p(L,t)$ at the subsurface. Let us time integrate the expression in Eq. (\ref{LimitDoublet}) with magnitutde $C(t)$ and take the limit $x \to L^{-}$ (i.e, observe the density at the subsurface due to a doublet of magnitude $C(t)$)

\begin{equation} \label{B:1}
\begin{array}{ll}
 \lim_{x \to L^{-}}\int_{0}^{t} C(\tau) \frac{L-x}{2\sqrt{\pi} \left [D (t-\tau)\right]^{3/2}} e^{-\frac{(L-x)^{2}}{4 D(t-\tau)}} d \tau
 \\
 =\lim_{x \to L^{-}}\frac{2}{D\sqrt{\pi}}\int_{\frac{L-x}{\sqrt{4 D t}}}^{\infty} C(t-\frac{(L-x)^2}{4 D \alpha^2}) e^{-\alpha^2} d \alpha
 \\
 =\frac{2C(t)}{D\sqrt{\pi}}\int_{0}^{\infty} e^{-\alpha^2} d \alpha
 \\
 =\frac{C(t)}{D},
\end{array}
\end{equation}

\noindent where we changed variables according to $\alpha=\frac{L-x}{\sqrt{4 D (t-\tau)}}$. Thus, to keep a subsurface density of $p(L,t)$ we set $C(t)= D p(L,t)$.


\renewcommand{\theequation}{C\arabic{equation}}
\renewcommand{\thefigure}{C\arabic{figure}}
\renewcommand{\thesection}{C\Roman{section}} 
\setcounter{equation}{0}
\section{Appendix C: Single-particle diffusing in a semi-infinite tube with an adsorbing surface}
\subsection{Initial position is in the bulk}

We assume the validity of Eqs. (\ref{A:1})-(\ref{A:5}), and we supplement these conditions with the linear dynamics closure relation (single-particle analogue of Eq. \ref{Henry}), whose Laplace transform is given by

\begin{equation}  \label{D:1}
      -D\frac{d \tilde{p}_{+}(x,s)}{d x}\big|_{x=L}  = k_{a} \tilde{p}_{+}(L,s) - k_{d} \tilde{\Pi}(s),
\end{equation}

\noindent where we have made use of Eq. (\ref{A:4}) to rewrite the left hand side.

The general form of the solution is in Eq. (\ref{A:6}). The above five conditions, when imposed on Eq. \ref{A:6}, produce a system of five equations with five unknowns ($\tilde{\Pi}(s)$ and $C_{i}(s)$ for $1 \leq i \leq 4$). Solving, we get 

\begin{equation}  \label{D:2}
\boxed{\tilde{\Pi}(s) = \frac{ K e^{-L\sqrt{\frac{s}{D}}}}{ K s + (\frac{s}{k_{d} }+1)\sqrt{D s}}  },
\end{equation}

\noindent where $K:=k_{a}/k_{d}$, and

\begin{equation}  \label{D:3}
\begin{cases}
   C_{1}(s) = \frac{\left(k_{d}+s\right) \operatorname{cosh}\left( L \sqrt{\frac{s}{D}}\right) + k_{a} \sqrt{\frac{s}{D}} \operatorname{sinh}\left( L \sqrt{\frac{s}{D}}\right)}{ \sqrt{D s^{3}}+k_{a} s+ k_{d} \sqrt{D s}}  e^{- L \sqrt{\frac{s}{D}}},
\\
    C_{2}(s) = 0  ,
\\
    C_{3}(s) = \frac{ \sqrt{D s^{3}}- k_{a} s + k_{d} \sqrt{D s} }{ 2s \left(k_{a} \sqrt{D s} + D (k_{d} + s)\right)} e^{- 2L \sqrt{\frac{s}{D}}},
\\
    C_{4}(s) = \frac{1}{2 \sqrt{D s}}.
\end{cases}
\end{equation}

\subsection{Initial position on the surface}
Here the boundary conditions Eqs. (\ref{A:3}) and (\ref{A:11}) are supplemented by the closure relation in Eq. (\ref{D:1}). The propagator takes the form of Eq. (\ref{A:10}).

The above three conditions, when imposed on Eq. (\ref{A:10}), produce a system of three equations with three unknowns ($\tilde{\Pi}(s)$ and $C_{i}(s)$ for $1 \leq i \leq 2$). Solving, we get 

\begin{equation}  \label{D:5}
\boxed{\tilde{\Pi}(s) =  \frac{K + \frac{\sqrt{D s}}{k_d} }{ K s + (\frac{s}{k_{d} }+1)\sqrt{D s}} },
\end{equation}

\noindent and

\begin{equation}  \label{D:6}
\begin{cases}
   C_{1}(s) =  \frac{ e^{-L\sqrt{\frac{s}{D}}} }{K s + (\frac{s}{k_{d} }+1)\sqrt{D s}},
\\
    C_{2}(s) = 0.  
\end{cases}
\end{equation}

\renewcommand{\theequation}{D\arabic{equation}}
\renewcommand{\thefigure}{D\arabic{figure}}
\renewcommand{\thesection}{D\Roman{section}} 
\setcounter{equation}{0}
\section{Appendix D: Inverse Laplace transform of Eq. (\ref{surface}) }

We hereby present two alternative approaches to the analytical inversion of Eq. (\ref{surface}). These yield two equivalent representations of the real-time behaviour, the second of which is identical to the solution presented by Agmon in the context of Smoluchowski-type reversible (association-dissociation) reactions \cite{agmon1984diffusion}.

\subsection{Inversion method A}

Let us first recall Eq. (\ref{surface}) from the main text

\begin{equation}  \label{E:1}
\tilde{\Pi}(s) =  \frac{ K e^{-L\sqrt{\frac{s}{D}}}}{ K s + (\frac{s}{k_{d} }+1)\sqrt{D s}},
\end{equation}

\noindent which we would like to invert back to the time domain. Equation (\ref{E:1}) can be written in  the following form

\begin{equation}  \label{E:2}
    f(s)=\frac{e^{-a\sqrt{s}}}{s+bs\sqrt{s}+c\sqrt{s}},
\end{equation}

\noindent where $a=\frac{L}{\sqrt{D}}$, $b=\frac{\sqrt{D}}{k_d K}=\frac{\sqrt{D}}{k_a}$ and $c=\frac{\sqrt{D}}{K}$. Thus, we are required to Laplace invert a function of the form (\ref{E:2}).

To do this, let us first rewrite Eq. (\ref{E:2})

\begin{align}
 \label{E:3}
      f(s) &=\frac{e^{-a\sqrt{s}}}{\sqrt{s}} \frac{c+bs-\sqrt{s}}{(c+bs)^2-s} 
      \nonumber \\
&=\frac{e^{-a\sqrt{s}}}{\sqrt{s}} \frac{1}{b^2}\left[ \frac{c+bs}{(s-s_1)(s-s_2)}- \frac{\sqrt{s}}{(s-s_1)(s-s_2)} \right] 
 \nonumber   \\
&= f_1(s)f_2(s)-\frac{1}{b^2}\frac{e^{-a\sqrt{s}}}{(s-s_1)(s-s_2)},
\end{align}

\noindent where $(c+bs)^2-s=b^2(s-s_1)(s-s_2)$ with $s_1=\frac{1-2bc+\sqrt{1-4bc}}{2b^2},~s_2=\frac{1-2bc-\sqrt{1-4bc}}{2b^2}$ and where $f_1(s)=\frac{e^{-a\sqrt{s}}}{\sqrt{s}}$ and $f_2(s)=\frac{c+bs}{(c+bs)^2-s}$. To invert the first term in Eq. (\ref{E:3}), we use the convolution property

\begin{equation}  \label{E:4}
    \mathcal{L}^{-1}_{s \to t} \left[ f_1(s)f_2(s) \right]=\int_0^t~dt'~f_1(t-t')~f_2(t'),
\end{equation}

\noindent where

\begin{equation}  \label{E:5}
 f_1(t)=\frac{e^{-a^2/4t}}{\sqrt{\pi t}},  
\end{equation}

\noindent and 

\begin{equation}  \label{E:6}
\begin{array}{ll}
   f_2(t)
    =\frac{1}{b^2} \Big(\frac{b s_1+c}{s_1-s_2} e^{s_1 t} +\frac{b s_2+c}{s_2-s_1} e^{s_2 t}\Big), 
\end{array}
\end{equation}

\noindent where we have used the well known formula for inversion of rational expressions \cite{spiegel1965schaum}. 

Plugging Eqs. (\ref{E:5}) and (\ref{E:6}) in Eq. (\ref{E:4}) and computing the convolution integral we obtain the first term of Eq. (\ref{E:3}):

\small\begin{align}  \label{E:7}
&      \mathcal{L}^{-1}_{s \to t} \left[ f_1(s)f_2(s)\right] = \frac{1}{2\sqrt{1-4bc}} \times
    \\& \nonumber
       \\& \nonumber
\Bigg\{ \left[ e^{s_1 t-a \sqrt{s_1}}\text{erfc}\left(\frac{a-2 \sqrt{s_1} t}{2 \sqrt{t}}\right)-e^{s_1 t+a \sqrt{s_1}} \text{erfc}\left(\frac{a+2 \sqrt{s_1} t}{2 \sqrt{t}}\right)\right]
\\& \nonumber
\\& \nonumber
+\left[e^{s_2 t+a \sqrt{s_2}} \text{erfc}\left(\frac{a+2 \sqrt{s_2} t}{2 \sqrt{t}}\right) -e^{s_2 t-a \sqrt{s_2}}\text{erfc}\left(\frac{a-2 \sqrt{s_2} t}{2
   \sqrt{t}}\right)\right]\Bigg\}
\end{align}\normalsize

As for the inversion of the second term in Eq. (\ref{E:3}), it is beneficial to rewrite it as follows

\begin{equation}  \label{E:8}
\begin{array}{ll}
  \frac{1}{b^2}\mathcal{L}^{-1}_{s \to t} \left[ \frac{e^{-a\sqrt{s}}}{(s-s_1)(s-s_2)}\right]
\\
\\
 = \frac{1}{\left(s_1-s_2\right)b^2}\mathcal{L}^{-1}_{s \to t} \left[\frac{e^{-a \sqrt{s}}}{s-s_1}-\frac{e^{-a
   \sqrt{s}}}{s-s_2}\right]
\\
\\
= \frac{1}{\left(s_1-s_2\right)b^2}\mathcal{L}^{-1}_{s \to t} \left[\frac{e^{-a \sqrt{s}}}{s}\frac{s}{s-s_1}-\frac{e^{-a
   \sqrt{s}}}{s}\frac{s}{s-s_2}\right].
\end{array}
\end{equation}

This expression is a sum of yet another two  convolution terms of the same form: a term $e^{-a \sqrt{s}}/s$ which has a well known inverse $\text{erfc}(a/2\sqrt{t})$ \cite{spiegel1965schaum}, and a term of the form $\frac{s}{s-\mu}$, where $\mu$ is either $s_1$ or $s_2$. We can invert the latter by using the following algebraic manipulation:

\begin{equation}  \label{E:9}
\begin{array}{ll}
\mathcal{L}^{-1}_{s \to t} \left[\frac{s}{s-\mu }\right]  =  \mathcal{L}^{-1}_{s \to t} \left[\frac{s-\mu+\mu}{s-\mu }\right]
= \mathcal{L}^{-1}_{s \to t} \left[1+\frac{\mu}{s-\mu}\right] 
\\
\\
=\mathcal{L}^{-1}_{s \to t}\left[1\right]+\mathcal{L}^{-1}_{s \to t}\left[\frac{\mu}{s-\mu}\right] = \delta(t) + \mu e^{\mu t}.
\end{array}
\end{equation}

\noindent After calculating the convolution integrals, Eq. (\ref{E:8}) reads

\begin{equation}  \label{E:10}
\begin{array}{ll}
  \frac{1}{b^2}\mathcal{L}^{-1}_{s \to t} \left[ \frac{e^{-a\sqrt{s}}}{(s-s_1)(s-s_2)}\right]
\\
\\
=\frac{e^{s_1 t-a \sqrt{s_1}} \left[\text{erfc}\left(\frac{a-2 \sqrt{s_1} t}{2 \sqrt{t}}\right)+e^{2 a \sqrt{s_1}} \text{erfc}\left(\frac{a+2 \sqrt{s_1} t}{2 \sqrt{t}}\right)\right]}{2\sqrt{1-4bc}}
 \\
 \\
-\frac{e^{s_2 t-a
   \sqrt{s_2}} \left[\text{erfc}\left(\frac{a-2 \sqrt{s_2} t}{2 \sqrt{t}}\right)+e^{2 a \sqrt{s_2}} \text{erfc}\left(\frac{a+2 \sqrt{s_2} t}{2 \sqrt{t}}\right)\right]}{2\sqrt{1-4bc}}.
\end{array}
\end{equation}

Subtracting the inversion in Eq. (\ref{E:10}) from (\ref{E:7}), we obtain the inversion of Eq. (\ref{E:1})  

\begin{align}  \label{E:11}
&     \Pi(t) = \frac{1}{\sqrt{1-4bc}} \times
    \\& \nonumber
       \\& \nonumber
\left[ e^{s_2 t+a \sqrt{s_2}} \text{erfc}\left(\frac{a+2 \sqrt{s_2} t}{2 \sqrt{t}}\right)-e^{s_1 t+a \sqrt{s_1}} \text{erfc}\left(\frac{a+2 \sqrt{s_1} t}{2 \sqrt{t}}\right)\right],
\end{align}\normalsize

\noindent where we recall that $a=\frac{L}{\sqrt{D}}$, $b=\frac{\sqrt{D}}{k_a}$ and $c=\frac{\sqrt{D}}{K}$ and $s_1=\frac{1-2bc+\sqrt{1-4bc}}{2b^2},~s_2=\frac{1-2bc-\sqrt{1-4bc}}{2b^2}$.

 \bigskip

\subsection{Inversion method B}

\noindent Equation (\ref{E:1}) can be written in  the following form

\begin{equation}  \label{EE:2}
    \tilde{\Pi}(s)=\frac{k_a e^{-L \sqrt{\frac{s}{D}}}}{\sqrt{s} \left(\sqrt{D} k_d+\sqrt{D} s+k_a \sqrt{s}\right)}.
\end{equation}
Using partial fraction decomposition the above equation can be written as
\begin{equation}  \label{EE:3}
    \tilde{\Pi}(s)=\frac{Ke^{-L \sqrt{\frac{s}{D}}}}{\sqrt{Ds}}-\frac{K\left(\sqrt{Ds}+k_a\right) e^{-L
   \sqrt{\frac{s}{D}}}}{\sqrt{D} \left(\sqrt{D} k_d+\sqrt{D} s+k_a \sqrt{s}\right)}.
\end{equation}

Going forward, we will refer to the first and second terms as $a_1, a_2$ respectively, such that $\tilde{\Pi}(s)=a_1-a_2$. The term $a_1$ can be easily inverted using a Laplace transform table to give $ K e^{-\frac{L^2}{4 D t}}/\sqrt{\pi D t}$. The term $a_2$ requires further simplification. We start by rewriting $a_2$ by factorizing its denominator

\begin{align}
\label{EE:4}
a_2 = \frac{K\left(k_a+\sqrt{Ds}\right) e^{-L \sqrt{\frac{s}{D}}}}{D \left(\sqrt{s}-\frac{1}{2} \sqrt{D}
   \left(-\frac{k_a}{D}-\Delta \right)\right) \left(\sqrt{s}-\frac{1}{2} \sqrt{D} \left(\Delta
   -\frac{k_a}{D}\right)\right)},
\end{align}
where $\Delta = \frac{1}{D}\sqrt{k_a^2-4 D k_d}$. Using partial fraction decomposition once more, we can further simplify the right hand side
\begin{align}
\label{EE:5}
a_2 = &\frac{Ke^{-L \sqrt{\frac{s}{D}}} \left(\sqrt{s}+\frac{k_a}{\sqrt{D}}\right)}{D \Delta  \left(\sqrt{s}+\frac{1}{2} \sqrt{D}
   \left(\frac{k_a}{D}-\Delta \right)\right)} \nonumber \\
   &-\frac{K e^{-L \sqrt{\frac{s}{D}}}
   \left(\sqrt{s}+\frac{k_a}{\sqrt{D}}\right)}{D \Delta   \left(\sqrt{s}+\frac{1}{2} \sqrt{D} \left(\frac{k_a}{D}+\Delta
   \right)\right)}.
\end{align}
Using a simple algebraic trick we can rewrite the fractions on the right hand side of the above equation as follows
\begin{align}
\label{EE:6}
a_2 = &\frac{Ke^{-L \sqrt{\frac{s}{D}}}}{D \Delta  }+\frac{Ke^{-L \sqrt{\frac{s}{D}}}}{D \Delta  }\frac{\frac{k_a}{2 \sqrt{D}}+\frac{\sqrt{D} \Delta }{2}}{\frac{1}{2} \sqrt{D}
   \left(\frac{k_a}{D}-\Delta \right)+\sqrt{s}} \nonumber \\
   &-\frac{Ke^{-L \sqrt{\frac{s}{D}}}}{D \Delta  }-\frac{Ke^{-L \sqrt{\frac{s}{D}}}}{D \Delta  }\frac{\frac{k_a}{2 \sqrt{D}}-\frac{\sqrt{D} \Delta }{2}}{\frac{1}{2} \sqrt{D}
   \left(\frac{k_a}{D}+\Delta \right)+\sqrt{s}}.
\end{align}
Denoting the terms on the right hand side of the above equation by $a_3, a_4, a_5, a_6$ (from left to write) we get
\begin{equation}  \label{EE:7}
\tilde{\Pi}(s) =  a_1 - a_2 = a_1 - a_3 - a_4 - a_5 - a_6.
\end{equation}

The above terms $a_1, a_3, a_4, a_5$ and $a_6$ can now be easily inverted using a Laplace transform table to give the following solution
\begin{align}  \label{EE:8}
\Pi(t) = &\frac{k_a \text{erfc}\left(\frac{t (k_a-D \Delta )+L}{2
   \sqrt{D t}}\right) e^{\frac{(k_a-D \Delta ) (t (k_a-D \Delta )+2 L)}{4 D}}}{D \Delta} \nonumber \\
   &-\frac{k_a \text{erfc}\left(\frac{t (D \Delta
   +k_a)+L}{2 \sqrt{D t}}\right) e^{\frac{(D \Delta +k_a) (t (D \Delta +k_a)+2 L)}{4
   D}}}{D\Delta},
\end{align}

\noindent which, under the correct re-scaling, is identical to the solution in \cite{agmon1984diffusion} (one minus Eq. (28) therein).

 \bigskip

\renewcommand{\theequation}{E\arabic{equation}}
\renewcommand{\thefigure}{E\arabic{figure}}
\renewcommand{\thesection}{E\Roman{section}} 
\setcounter{equation}{0}
\begin{widetext}
\section{Appendix E: Single-particle diffusing in a closed tube with two identical adsorbing surfaces}
\subsection{Initial position is in the bulk}

By the exact same method employed in appendices A and C, but this time in accordance with the conditions stated in Eqs. (\ref{box1}) and (\ref{box2}), we solve for $\tilde{\Pi}_{\pm}(s)$ and $C_{i}(s)$ ($1 \leq i \leq 4$):

\begin{equation} \label{F:1}
\begin{cases}
 \tilde{\Pi}_{+}(s)=   \frac{k_a \left(-s k_a+k_d \sqrt{D s}+\sqrt{D s^3}\right) \cosh \left(\sqrt{\frac{s}{D}}
   \left(L+x_0\right)\right)+s k_a^2 e^{\sqrt{\frac{s}{D}} \left(L+x_0\right)}}{2 s k_a \left(k_d \sqrt{D
   s}+\sqrt{D s^3}\right) \cosh \left(2 L \sqrt{\frac{s}{D}}\right)+s \left(s k_a^2+D \left(k_d+s\right){}^2\right) \sinh \left(2 L
   \sqrt{\frac{s}{D}}\right)}
 \\
 \\
 \tilde{\Pi}_{-}(s)=   \frac{e^{\sqrt{\frac{s}{D}}\left(L-x_{0}\right)} \left[\left(e^{2 L \sqrt{\frac{s}{D}}}-e^{2 \sqrt{\frac{s}{D}} x_{0}}\right)k_{a} s+\sqrt{D}\left(e^{2 L \sqrt{\frac{s}{D}}}+e^{2 \sqrt{\frac{s}{D}} x_{0}}\right)\left(s^{3 / 2}+k_{d} \sqrt{s}\right)\right]}{s\left[\left(e^{4 L \sqrt{\frac{s}{D}}}-1\right)  k_{a}s+\frac{D}{k_{a}}\left(e^{4 L \sqrt{\frac{s}{D}}}-1\right)\left(s+k_{d}\right)^{2}+2\left(e^{4 L \sqrt{\frac{s}{D}}}+1\right) \sqrt{D}\left(s^{3 / 2}+k_{d} \sqrt{s}\right)\right]}
\end{cases},
\end{equation}

\noindent and

\begin{equation} \label{F:2}
\begin{cases}
C_1(s)=\frac{e^{-\sqrt{\frac{s}{D}}\left(2 L+x_{0}\right)}\left(\sqrt{D s^{3}}-k_{a}s + k_{d}\sqrt{D s}\right)\left[\sqrt{D s^{3}}- k_{a}s+e^{2 \sqrt{\frac{s}{D}}\left(L+x_{0}\right)} \left(\sqrt{D s}+k_{a}\right)s+k_{d} \sqrt{D s}\left(e^{2 \sqrt{\frac{s}{D}}\left(L+x_{0}\right)}+1\right)\right]}{4 \sqrt{D} s^{3 / 2}\left[2 k_{a} \sqrt{D s} \cosh \left(2 L \sqrt{\frac{s}{D}}\right)\left(s+k_{d}\right)+ \sinh \left(2 L \sqrt{\frac{s}{D}}\right) \left[D\left(s+k_{d}\right)^{2}+k_{a}^{2} s\right]\right]},
\\
\\
C_2(s)=\frac{e^{-\sqrt{\frac{s}{D}} x_{0}}\left(\sqrt{D s^{3}}+k_{a} s+k_{d} \sqrt{D s}\right)\left[\sqrt{D} s^{3 / 2}\left(e^{2 \sqrt{\frac{s}{D}}\left(L+x_{0}\right)}+1\right)+\left(e^{2 \sqrt{\frac{s}{D}}\left(L+x_{0}\right)}-1\right) k_{a} s+\left(e^{2 \sqrt{\frac{s}{D}}\left(L+x_{0}\right)}+1\right) k_{d} \sqrt{D s}\right]}{4 \sqrt{D} s^{3 / 2}\left[2 k_{a} \sqrt{D s} \cosh \left(2 L \sqrt{\frac{s}{D}}\right)\left(s+k_{d}\right)+ \sinh \left(2 L \sqrt{\frac{s}{D}}\right) \left[D\left(s+k_{d}\right)^{2}+k_{a}^{2} s\right]\right]},
\\
\\
C_3(s)=\frac{se^{-\sqrt{\frac{s}{D}} x_{0}}\left[k_{a}^{2}\left(e^{2 L \sqrt{\frac{s}{D}}}-e^{2 \sqrt{\frac{s}{D}} x_{0}}\right)  s+2 k_{a} \sqrt{D s} e^{2 L \sqrt{\frac{s}{D}}}\left(s+k_{d} \right)+D\left(e^{2 L \sqrt{\frac{s}{D}}}+e^{2 \sqrt{\frac{s}{D}} x_{0}}\right)\left(s+k_{d}\right)^{2}\right]}{4 \sqrt{D} s^{3 / 2}\left[2 k_{a} \sqrt{D s} \cosh \left(2 L \sqrt{\frac{s}{D}}\right)\left(s+k_{d}\right)+ \sinh \left(2 L \sqrt{\frac{s}{D}}\right) \left[D\left(s+k_{d}\right)^{2}+k_{a}^{2} s\right]\right]},
\\
\\
C_4(s)=\frac{-se^{-\sqrt{\frac{s}{D}}\left(2 L+x_{0}\right)}\left[k_{a}^{2}\left(e^{2 L \sqrt{\frac{s}{D}}}-e^{2 \sqrt{\frac{s}{D}} x_{0}}\right) s+2 k_{a} \sqrt{D s} e^{2 \sqrt{\frac{s}{D}} x_{0}}\left(s+k_{d}\right)-D\left(e^{2 L \sqrt{\frac{s}{D}}}+e^{2 \sqrt{\frac{s}{D}} x_{0}}\right)\left(s+k_{d}\right)^{2}\right]}{4 \sqrt{D} s^{3 / 2}\left[2 k_{a} \sqrt{D s} \cosh \left(2 L \sqrt{\frac{s}{D}}\right)\left(s+k_{d}\right)+ \sinh \left(2 L \sqrt{\frac{s}{D}}\right) \left[D\left(s+k_{d}\right)^{2}+k_{a}^{2} s\right]\right]}.
\end{cases}
\end{equation}
\end{widetext}

Taking the limit $k_d \to 0$ we have

\begin{align} \label{F:3}
 \tilde{\Pi}_{\pm}(s) &= \\
 &\frac{\sqrt{D s} \cosh \left[\sqrt{\frac{s}{D}}\left(L\pm x_{0}\right)\right]+k_{a} \sinh \left[\sqrt{\frac{s}{D}}\left(L\pm x_{0}\right)\right]}{2 \sqrt{D} s^{3 / 2} \cosh \left[2 L \sqrt{\frac{s}{D}}\right]+s^{2}\left(\frac{k_{a}}{s}+\frac{D}{k_{a}}\right) \sinh \left[2 L \sqrt{\frac{s}{D}}\right]}. \nonumber
\end{align}

Similarly, taking the limit $k_a \to \infty$ we have

\begin{equation} \label{F:4}
 \tilde{\Pi}_{\pm}(s) =
\frac{e^{\sqrt{\frac{s}{D}}\left(3 L\pm x_{0}\right)}-e^{\sqrt{\frac{s}{D}}\left(L \mp x_{0}\right)}}{\left(e^{4 L \sqrt{\frac{s}{D}}}-1\right) s},
\end{equation}

\noindent and, of course, when $k_a \to 0$ or $k_d \to \infty$ we have $\Pi_{\pm}(t)=0$.

\renewcommand{\theequation}{F\arabic{equation}}
\renewcommand{\thefigure}{F\arabic{figure}}
\renewcommand{\thesection}{F\Roman{section}} 
\setcounter{equation}{0}

\section{Appendix F: Uniform initial concentration in a closed tube with two identical adsorbing surfaces} \setcounter{equation}{0}

The boundary conditions for this problem are identical to these of the single-prticle problem, and so are their Laplace transforms. The initial condition, however, is different. For all $-L<x<L$ we have:

\begin{equation} \label{G:1}
c(x,0)=c_{b}.    
\end{equation}

\noindent Thus, the Laplace transformation of the 
diffusion equation results in

\begin{equation} \label{G:2}
   s \tilde{c}(x,s) - c_b = D\frac{d^2 \tilde{c}(x,s)}{dx^2},
\end{equation}

\noindent which is a second-order, linear, \textit{non-homogeneous} differential equation.  

As before, the homogeneous solution for this equation can be written as

\begin{equation}  \label{G:3}
\tilde{c}_{H}(x,s)= C_{1}(s) e^{x\sqrt{\frac{s}{D}}} + C_{2}(s) e^{-x\sqrt{\frac{s}{D}}},
\end{equation}

\noindent and it is easy to the verify that

\begin{equation}  \label{G:4}
\tilde{c}_{P}(x,s)= \frac{c_b}{s},
\end{equation}

\noindent is a particular solution. The overall solution is given by $\tilde{c}_{H}(x,s)+\tilde{c}_{P}(x,s)$. Imposing the conditions one gets: 

\begin{equation}  \label{G:5}
\tilde{\Gamma}(s) = \frac{2 k_{a}\operatorname{sinh}( L \sqrt{\frac{s}{D}})}{\left(s^{2}+k_{d} s \right)\operatorname{sinh}( L \sqrt{\frac{s}{D}})+\frac{s^{3 / 2}}{\sqrt{D}} k_{a}\operatorname{cosh}( L \sqrt{\frac{s}{D}})} c_{b}.
\end{equation}

\noindent and

\begin{equation}  \label{G:6}
\begin{array}{ll}
C_1(s)=C_2(s)   
\\
=-\frac{1}{2}\frac{k_{a} c_{b}}{ k_{a} s \cosh \left(L \sqrt{\frac{s}{D}}\right)+ \sqrt{D s}\left(s+k_{d}\right) \sinh \left(L \sqrt{\frac{s}{D}}\right)}.       
\end{array}
\end{equation}

\renewcommand{\theequation}{G\arabic{equation}}
\renewcommand{\thefigure}{G\arabic{figure}}
\renewcommand{\thesection}{G\Roman{section}} 
\section{Appendix G: Derivation of the Generalized Ward-Tordai Relation by the Green's Function Method} \setcounter{equation}{0}

Diffusion in a medium with an adsorbing surface can be modeled by the diffusion equation with non-homogeneous Dirichlet boundary conditions. More specifically, let $\mathcal{D}$ be a three-dimensional domain bounded by a closed surface $S$ (closed in the broad sense). Let $p(\Vec{r},t)$ be the probability density function to find the diffusing particle at $\Vec{r}$ at time $t$, given an initial concentration profile $p_{0}(\Vec{r})$ and a time-dependent subsurface probability density $\phi(\Vec{r},t)$. The subsurface density is an unspecified non-homogeneous boundary condition. The probability to find the adsorbate in the bulk at time $t$ is then simply given by $\iiint_{\mathcal{D}} p(\Vec{r},t) d\Vec{r}$. If the particle is not in the bulk, it will surely be adsorbed to the surface, and thus one has 
\begin{align} \label{G:inline}
\Pi(t) = 1 - \iiint_{\mathcal{D}} p(\Vec{r},t) d\Vec{r}.    
\end{align}

\noindent To find $p(\Vec{r},t)$ we closely follow \cite{carslaw1959conduction}, with modified notation to fit our discussion here, and additional comments to clarify the calculations made. In particular, we employ the Green's function method to solve the problem.

In our set-up, the Green's function, denoted by $g(\Vec{r},t \mid \Vec{r}',\tau)$, is the probability to find a diffusing single-particle at $\Vec{r}$ at time $t$, given that it was instantaneously introduced to the domain $\mathcal{D}$ at $\Vec{r}'$ at time $\tau$ exactly (delta function initial condition, $\delta(\Vec{r}-\Vec{r}')$, also known as an instantaneous source). The surface $S$ is unchanged, but recall that the Green's function is defined for the case where it is \textit{absorbing}. Solving for the Green's function in this set-up is a considerably simpler task, and in the following we will show how to express the solution for the original problem in terms of this Green's function.
    
Since the particle is diffusing, $g(\Vec{r},t \mid \Vec{r}',\tau)$ follows the ordinary diffusion equation for all $t>\tau$:
    \begin{equation} \label{C:1}
        \frac{\partial g(\Vec{r},t \mid \Vec{r}',\tau)}{\partial t}=D \nabla^{2} g(\Vec{r},t \mid \Vec{r}',\tau),
\end{equation}

\noindent where $\nabla$ is the Laplace operator with respect to $\Vec{r}$. Equation (\ref{C:1}) is a forward Fokker-Planck equation, the corresponding backward Fokker-Planck equation is (See pp. 47-48 in Ref. \cite{paul1999stochastic})

\begin{equation} \label{C:2}
        \frac{\partial g(\Vec{r},t \mid \Vec{r}',\tau)}{\partial \tau}=-D \nabla'^{2} g(\Vec{r},t \mid \Vec{r}',\tau),
\end{equation}

\noindent where $\nabla'$ is the Laplace operator with respect to $\Vec{r}'$. Essentially, this is a partial differential equation where the initial parameters $(\Vec{r}',\tau)$ now play the role of variables \cite{paul1999stochastic}. 

To proceed further, we note the following
\begin{align}  \label{C:5}
&\frac{\partial}{\partial \tau}\left[g(\Vec{r},t \mid \Vec{r}',\tau) p(\Vec{r}',\tau)\right]
\\
&=g(\Vec{r},t \mid \Vec{r}',\tau) \frac{\partial p(\Vec{r}',\tau)}{\partial \tau}+p(\Vec{r}',\tau) \frac{\partial g(\Vec{r},t \mid \Vec{r}',\tau)}{\partial \tau} \nonumber
\\
&=D\left[g(\Vec{r},t \mid \Vec{r}',\tau) \nabla'^{2} p(\Vec{r}',\tau)-p(\Vec{r}',\tau) \nabla'^{2} g(\Vec{r},t \mid \Vec{r}',\tau)\right], \nonumber
\end{align}
\noindent where going from the second to the third line, we have used Eq. (\ref{C:2}) and further used the fact that $p(\Vec{r},t)$ obeys the diffusion equation $\frac{\partial p(\Vec{r}',\tau)}{\partial \tau}=D \nabla'^{2} p(\Vec{r}',\tau),$ regardless of the notation chosen for the space and time variables. Indeed, unlike Eq. (\ref{C:2}), the latter is a simple forward diffusion equation for $p(\Vec{r}',\tau)$. To further clarify, while $(\Vec{r}',\tau)$ are considered to be the final variables for $p(\Vec{r}',\tau),$ they are the initial variables for $g(\Vec{r},t \mid \Vec{r}',\tau)$.

Integrating both sides of  Eq. (\ref{C:5}) over all the domain with respect to $\Vec{r}'$ and over $\tau$ from $0$ to some time $t-\epsilon$, where $0<\epsilon\ll t$, we have 

\begin{equation}  \label{C:6}
\begin{array}{ll} 
\int_{0}^{t-\epsilon}\iiint_{\mathcal{D}}\frac{\partial}{\partial \tau}\left[g(\Vec{r},t \mid \Vec{r}',\tau)  p(\Vec{r}',\tau)\right] d \Vec{r}' d \tau
\\
\\
=D\int_{0}^{t-\epsilon}\iiint_{\mathcal{D}}\Big[g(\Vec{r},t \mid \Vec{r}',\tau) \nabla'^{2} p(\Vec{r}',\tau)
\\
\\
-p(\Vec{r}',\tau) \nabla'^{2} g(\Vec{r},t \mid \Vec{r}',\tau)\Big] d \Vec{r}' d \tau.
\end{array}
\end{equation}

\noindent On the left-hand side we interchange the order of integration and time-integrate to get

\begin{equation}  \label{C:7}
\begin{array}{ll} 
\iiint_{\mathcal{D}}\int_{0}^{t-\epsilon}\frac{\partial}{\partial \tau}\left[g(\Vec{r},t \mid \Vec{r}',\tau)  p(\Vec{r}',\tau)\right] d \tau d \Vec{r}'
\\
\\
=\iiint_{\mathcal{D}}g(\Vec{r},t \mid \Vec{r}',t-\epsilon)  p(\Vec{r}',t-\epsilon) d \Vec{r}' 
\\
\\
- \iiint_{\mathcal{D}}g(\Vec{r},t \mid \Vec{r}',0)  p_{0}(\Vec{r}') d \Vec{r}',
\end{array}
\end{equation}

\noindent where we have used the initial condition $ p(\Vec{r}',0) \equiv p_{0}(\Vec{r}')$. By taking the limit $\epsilon \to 0$ we have $g(\Vec{r},t \mid \Vec{r}',t-\epsilon) \to \delta(\Vec{r}-\Vec{r}')$, and Eq. (\ref{C:7}) becomes

\begin{equation}  \label{C:8}
\begin{array}{ll} \lim_{\epsilon \to 0}
\iiint_{\mathcal{D}}\int_{0}^{t-\epsilon}\frac{\partial}{\partial \tau}\left[g(\Vec{r},t \mid \Vec{r}',\tau)  p(\Vec{r}',\tau)\right] d \Vec{r} d \tau
\\
\\
=p(\Vec{r},t)
- \iiint_{\mathcal{D}}g(\Vec{r},t \mid \Vec{r}',0)  p_{0}(\Vec{r}') d \Vec{r}'.
\end{array}
\end{equation}

We continue to evaluate the right-hand side of Eq. (\ref{C:6}). To this end,
we employ Green's second identity, which is essentially the Gauss's Divergence Theorem applied on the divergence of a multiplication of a scalar and a vector field (in our case we have the Laplace operator, which is the divergence of the gradient, where the latter is indeed a vector field). Thus the right-hand side reads (after taking the limit $\epsilon \to 0$)

\begin{equation}  \label{C:9}
\begin{array}{ll} 
D\int_{0}^{t}\iiint_{\mathcal{D}}\Big[g(\Vec{r},t \mid \Vec{r}',\tau) \nabla'^{2} p(\Vec{r}',\tau)
\\
\\
-p(\Vec{r}',\tau) \nabla'^{2} g(\Vec{r},t \mid \Vec{r}',\tau)\Big] d \Vec{r}' d \tau
\\
\\
=D \int_{0}^{t}\Big[\oiint_{S}\big[-g(\Vec{r},t \mid \Vec{r}',\tau) \frac{\partial  p(\Vec{r}',\tau)}{\partial \Vec{n}'_{i}}
\\
\\
+ p(\Vec{r}',\tau) \frac{\partial g(\Vec{r},t \mid \Vec{r}',\tau)}{\partial \Vec{n}'_{i}}\big] d \Vec{r}'\Big] d \tau,
\end{array}
\end{equation}

\noindent where  $\frac{\partial}{\partial \Vec{n}'_{i}}$ denotes differentiation along the inward-drawn normal and the volume integral has now become a surface integral. Note that in moving from the outward- to the inward-drawn normal one multiplies by $-1$.  Recall that, the Green's function is, by definition, the solution to the problem with an absorbing surface $S$, and so $g(\Vec{r},t \mid \Vec{r}',\tau)=0$ for every $\Vec{r} \in S$ and the first term in the spatial integral in Eq. (\ref{C:9}) always contributes zero.

Collecting all the terms in Eq. (\ref{C:9}) and plugging them back to Eq. (\ref{C:6}), we have

\begin{equation}  \label{C:10}
\begin{array}{ll} 
p(\Vec{r},t)=\iiint_{\mathcal{D}}g(\Vec{r},t \mid \Vec{r}',0)  p_{0}(\Vec{r}') d \Vec{r}'
\\
\\
+D \int_{0}^{t}\left[\oiint_{S}p(\Vec{r}',\tau)\frac{\partial g(\Vec{r},t \mid \Vec{r}',\tau)} {\partial\Vec{n}'_{i}} d \Vec{r}'\right] d \tau,
\end{array}
\end{equation}

\noindent where we have also used Eq. (\ref{C:8}). Recall now that for the adsorbing surface problem we want to compute $\Pi(t)$ according to Eq. (\ref{G:inline}). Substituting $p(\Vec{r},t)$ from Eq. (\ref{C:10}) in the above definition, we obtain

\begin{equation}  \label{C:11}
\begin{array}{ll} 
\Pi(t) = 1 - \iiint_{\mathcal{D}} \left[\iiint_{\mathcal{D}}g(\Vec{r},t \mid \Vec{r}',0) d \Vec{r}\right]  p_{0}(\Vec{r}') d \Vec{r'}
\\
\\
-D \int_{0}^{t}\left(\oiint_{S}p(\Vec{r}',\tau) \left[\iiint_{\mathcal{D}}\frac{\partial g(\Vec{r},t \mid \Vec{r}',\tau)}{\partial \Vec{n}'_{i}}d \Vec{r}\right] d \Vec{r}'\right) d \tau.
\end{array}
\end{equation}

Note that the double triple integral in the above equation is simply the probability of finding the particle in the bulk at time $t$ given its initial location distribution $p_0(\Vec{r})$. In other words, this is the survival probability that is defined in the main text

\begin{align}
Q(t)=    \iiint_{\mathcal{D}} \left[\iiint_{\mathcal{D}}g(\Vec{r},t \mid \Vec{r}',0) p_{0}(\Vec{r}') d \Vec{r'}\right]   d \Vec{r}.
\end{align}

Also, $F(t)=1-Q(t)$, where $F(t)$ is the cumulative of the first passage time to the boundary as was defined in the main text. 

The derivation used here is quite generic. The 1- and 2-dimensional analogues relations are easily attainable by following the above steps with suitable modifications.
It can be appreciated that Eq. (\ref{C:11}) can indeed be obtained from the general relation in Eq. (\ref{GeneralizedWardTordai}) by assuming a 3-dimensional space.

 \bigskip

\renewcommand{\theequation}{H\arabic{equation}}
\renewcommand{\thefigure}{H\arabic{figure}}
\renewcommand{\thesection}{H\Roman{section}} 
\setcounter{equation}{0}
\begin{widetext} 
\section{Appendix H: Diffusion inside an adsorbing spherical shell}

Imposing the Laplace transform of the conditions in Eqs. (\ref{radial2}) and (\ref{radial3}), along with a linear dynamics closure relation and matching conditions on Eq. (\ref{radial5}) we solved for $\tilde{\Pi}_{\pm}(s)$ and $A_1$, $A_2$, $B_1$ and $B_2$. Because these terms are very cumbersome, we give here only $\tilde{\Pi}_{\pm}(s)$ which was used to get the results in the main text:

\vspace{1ex}

\begin{align} \label{H:1}
&\tilde{\Pi}_{+}(s)=   \nonumber \\
&\left(k_{a}\left(  (-s^{3 / 2} \sqrt{D} R_{-}-k_{d} R_{-}\sqrt{s D}) \cosh \left[\sqrt{\frac{s}{D}}\left(r_{0}-R_{-}\right)\right] -\sinh \left[\sqrt{\frac{s}{D}}\left(r_{0}-R_{-}\right)\right]\left(D\left(s+k_{d}\right)+s k_{a} R_{-}\right)\right) R_{+}^{2}\right)   \nonumber \\
&\left\{r _ { 0 } \left(2(s D)^{3 / 2} \cosh \left[\sqrt{\frac{s}{D}}\left(R_{-}-R_{+}\right)\right] k_{d} R_{-}-2(s D)^{3 / 2} \cosh \left[\sqrt{\frac{s}{D}}\left(R_{-}-R_{+}\right)\right] k_{d} R_{+}+\right.\right. \nonumber \\ 
&\quad \sqrt{s D} \cosh \left[\sqrt{\frac{s}{D}}\left(R_{-}-R_{+}\right)\right]\left(D\left(s^{2}+k_{d}^{2}\right) R_{-}-\left(D\left(s^{2}+k_{d}^{2}\right)+2 s k_{a}\left(s+k_{d}\right) R_{-}\right) R_{+}\right)+ \\ 
&\left.\left.\quad \sinh \left[\sqrt{\frac{s}{D}}\left(R_{-}-R_{+}\right)\right]\left(-D\left(s+k_{d}\right)\left(D\left(s+k_{d}\right)+s k_{a} R_{-}\right)+s\left(D k_{a}\left(s+k_{d}\right)+\left(s k_{a}^{2}+D\left(s+k_{d}\right)^{2}\right) R_{-}\right) R_{+}\right)\right)\right\}^{-1}, \nonumber
\end{align}

\noindent and 

\begin{align} \label{H:2}
 &\tilde{\Pi}_{-}(s)=\nonumber\\ 
 &\left(k_a R_{-}^2 \sinh \left(\left(r_0-R_{+}\right) \sqrt{\frac{s}{D }}\right) \left(-s k_a R_{+}
   \sqrt{\frac{s}{D }}+k_d \sqrt{D  s}+\sqrt{D  s^3}\right)+s k_a R_{-}^2 R_{+}
   \left(k_d+s\right) \cosh \left(\left(r_0-R_{+}\right) \sqrt{\frac{s}{D }}\right)  \right) \\ \nonumber
   &\Bigg\{ r_0 \sqrt{\frac{s}{D }} \cosh \left(\sqrt{\frac{s}{D }} \left(R_{-}-R_{+}\right)\right) \Bigg(2 k_a
   R_{-} R_{+} k_d \sqrt{D  s^3}+2 k_a R_{-} R_{+} \sqrt{D  s^5}-R_{-} k_d^2 \sqrt{D ^3 s}-2 R_{-} k_d
   (D  s)^{3/2}- \\ \nonumber
   & D ^{3/2} s^{5/2} R_{-}+R_{+} k_d^2 \sqrt{D ^3 s}+ 2 R_{+} k_d (D  s)^{3/2}+s
   R_{+} (D  s)^{3/2}\Bigg)+r_0 \sqrt{\frac{s}{D }} \sinh \left(\sqrt{\frac{s}{D }}
   \left(R_{-}-R_{+}\right)\right) \\ \nonumber
   & \Bigg(D  \left(k_d+s\right) \left(s k_a R_{-}+D 
   \left(k_d+s\right)\right) -s R_{+} \left(R_{-} \left(s k_a^2+D  \left(k_d+s\right){}^2\right)+D 
   k_a \left(k_d+s\right)\right)\Bigg) \Bigg\}^{-1}.
\end{align}

\end{widetext}


\renewcommand{\theequation}{I\arabic{equation}}
\renewcommand{\thefigure}{I\arabic{figure}}
\renewcommand{\thesection}{I\Roman{section}} 
\setcounter{equation}{0}
\section{Appendix I: Derivation of Eq. (\ref{outsidesphere4})}

Taking the $s \to 0$ limit of Eq. (\ref{outsidesphere1}) we obtain 

\begin{equation} \label{I:1}
     \tilde{\Pi}_{-}(s) \simeq \Lambda_\infty -  \frac{K R_{-}^{2}}{D^{3/2}}\sqrt{s},
\end{equation}

\noindent where $\Lambda_\infty$ for this case was calculated in Eq. (\ref{outsidesphere3}). 

Inversion of Eq. (\ref{I:1}) can be done in the following indirect manner. Let us define 

\begin{equation} \label{I:2}
     \Psi (t) = \int_t^{\infty} \Pi_{-}(t') dt'.
\end{equation}

\noindent It is a simple calculation to show that the Laplace transform of $\Psi (t)$ is 

\begin{equation} \label{I:3}
    \tilde{\Psi}(s) = \frac{\Lambda_\infty - \Pi(s)}{s},
\end{equation}

\noindent and plugging in Eqs. (\ref{outsidesphere3}) and (\ref{I:1}) gives

\begin{equation} \label{I:4}
    \tilde{\Psi}(s) \simeq \frac{K R_{-}^{2}}{D^{3/2}} s^{-1/2},
\end{equation}

\noindent in the $s \to 0$ limit.
Equation (\ref{I:4}) can be inverted using the Tauberian theorem which gives

\begin{equation} \label{I:5}
     \Psi(t) \simeq \frac{K R_{-}^{2}}{\sqrt{\pi D^{3}} } t^{-1/2},
\end{equation}

\noindent for $t \to \infty$.

Now note that by evoking the definition of $\Lambda(t)$ in Eq. (\ref{name}), another way to write Eq. (\ref{I:2}) is 

\begin{equation} \label{I:6}
     \Psi (t) = \Lambda_\infty - \Lambda(t).
\end{equation}

\noindent Hence,

\begin{equation} \label{I:7}
  \Pi_{-}(t) = - \frac{d \Psi (t)}{d t}.
\end{equation}

\noindent Plugging (\ref{I:5}) into (\ref{I:7}) we obtain the desired inversion of Eq. (\ref{I:1}):

\begin{equation} \label{I:8}
  \Pi_{-}(t) \simeq \frac{K R_-^2}{\sqrt{4 \pi D^3}}t^{-3/2},
\end{equation}

\noindent which applies for $t \to \infty$.

{}




\begin{thebibliography}{}

\bibitem{langmuir1916constitution}
I. Langmuir (1916). The constitution and fundamental properties of solids and liquids. Part I. Solids. \textit{Journal of the American chemical society}, 38(11), 2221-2295.

\bibitem{langmuir1917constitution}
I. Langmuir (1917). The constitution and fundamental properties of solids and liquids. II. Liquids. \textit{Journal of the American chemical society}, 39(9), 1848-1906.

\bibitem{langmuir1918adsorption}
I. Langmuir (1918). The adsorption of gases on plane surfaces of glass, mica and platinum. \textit{Journal of the American Chemical society}, 40(9), 1361-1403.

\bibitem{perrin2013brownian}
J. Perrin (2013). \textit{Brownian movement and molecular reality}, (Courier Corporation,2013).

\bibitem{einstein1905molekularkinetischen}
A. Einstein (1905). Über die von der molekularkinetischen Theorie der Wärme geforderte Bewegung von in ruhenden Flüssigkeiten suspendierten Teilchen. \textit{Annalen der physik}, 4.


\bibitem{elenko2009single}
M.P. Elenko, J.W. Szostak, A.M. van Oijen (2009). Single-molecule imaging of an in vitro-evolved RNA aptamer reveals homogeneous ligand binding kinetics. \textit{Journal of the American Chemical Society}, 131(29), 9866-9867.

\bibitem{elenko2010single}
M.P. Elenko, J.W. Szostak, A.M. van Oijen (2010). Single-molecule binding experiments on long time scales. \textit{Review of Scientific Instruments}, 81(8), 083705.

\bibitem{van2011single}
A.M. Van Oijen (2011). Single-molecule approaches to characterizing kinetics of biomolecular interactions. \textit{Current opinion in biotechnology}, 22(1), 75-80.

\bibitem{van2014single}
M.H. Van Es, J. Tang, J. Preiner, P. Hinterdorfer, T.H. Oosterkamp (2014). Single molecule binding dynamics measured with atomic force microscopy. \textit{Ultramicroscopy}, 140, 32-36.

\bibitem{gong2014direct}
X. Gong, Z. Wang, T. Ngai (2014). Direct measurements of particle–surface interactions in aqueous solutions with total internal reflection microscopy. \textit{Chemical Communications}, 50(50), 6556-6570.

\bibitem{shen2016single}
H. Shen, L.J. Tauzin, W. Wang, B. Hoener, B. Shuang, L. Kisley, A. Hoggard, C.F. Landes (2016). Single-molecule kinetics of protein adsorption on thin nylon-6, 6 films. \textit{Analytical chemistry}, 88(20), 9926-9933.

\bibitem{wang2017three}
D. Wang, H. Wu, D.K. Schwartz (2017). Three-dimensional tracking of interfacial hopping diffusion. \textit{Physical Review Letters}, 119(26), 268001.

\bibitem{borberg2019light}
E. Borberg, M. Zverzhinetsky, A. Krivitsky, A. Kosloff, O. Heifler, G. Degabli, H. Peretz-Soroka, R. Satchi-Fainaro, L. Burstein, S. Reuveni, H. Diamant, V. Krivitsky and F. Patolsky (2019). Light-controlled selective collection-and-release of biomolecules by an on-chip nanostructured device. \textit{Nano letters}, 19(9), 5868-5878.

\bibitem{wang2019probing}
H. Wang, Z. Tang, Y. Wang, G. Ma, N. Tao (2019). Probing single molecule binding and free energy profile with plasmonic imaging of nanoparticles. \textit{Journal of the American Chemical Society}, 141(40), 16071-16078.

\bibitem{wang2020probing}
Y. Wang, W. Jing, N. Tao, H. Wang (2020). Probing single-molecule binding event by the dynamic counting and mapping of individual nanoparticles. \textit{ACS sensors}, 6(2), 523-529.

\bibitem{wang2020non}
D. Wang, D.K. Schwartz (2020). Non-Brownian Interfacial Diffusion: Flying, Hopping, and Crawling. \textit{The Journal of Physical Chemistry C}, 124(37), 19880-19891.

\bibitem{Betzig2016}
E. Betzig, G.H. Patterson, R. Sougrat, O.W. Lindwasser, S. Olenych, J.S. Bonifacino, M.W. Davidson, J. Lippincott-Schwartz, H.F. Hess (2006). Imaging intracellular fluorescent proteins at nanometer resolution. \textit{Science}, 313(5793), 1642-1645.

\bibitem{Hell1994}
S.W. Hell, J. Wichmann (1994). Breaking the diffraction resolution limit by stimulated emission: stimulated-emission-depletion fluorescence microscopy. \textit{Optics Letters}, 19(11), 780-782.

\bibitem{Moerner1997}
R.M. Dickson, A.B. Cubitt, R.Y. Tsien, W.E. Moerner (1997). On/off blinking and switching behaviour of single molecules of green fluorescent protein. \textit{Nature}, 388(6640), 355-358.

\bibitem{Xie1998}
H.P. Lu, L. Xun, X.S. Xie (1998). Single-molecule enzymatic dynamics. \textit{Science}, 282(5395), 1877-1882.

\bibitem{Block1999}
K. Visscher, M.J. Schnitzer, S.M. Block (1999). Single kinesin molecules studied with a molecular force clamp. \textit{Nature}, 400(6740), 184-189.

\bibitem{Bustamante2003}
C. Bustamante, Z. Bryant S.B. Smith (2003). Ten years of tension: single-molecule DNA mechanics. \textit{Nature}, 421(6921), 423-427.

\bibitem{Gaub1997}
M. Rief, M. Gautel, F. Oesterhelt, J.M. Fernandez, H.E. Gaub (1997). Reversible unfolding of individual titin immunoglobulin domains by AFM. \textit{Science}, 276(5315), pp.1109-1112.

\bibitem{adamczyk1987adsorption}
Z. Adamczyk, J. Petlicki (1987). Adsorption and desorption kinetics of molecules and colloidal particles. \textit{Journal of colloid and interface science}, 118(1), 20-49.

\bibitem{baret1968kinetics}
J.F. Baret (1968). Kinetics of adsorption from a solution. Role of the diffusion and of the adsorption-desorption antagonism. \textit{The Journal of Physical Chemistry}, 72(8), 2755-2758.

\bibitem{chattoraj1984miscellaneous}
D.K. Chattoraj, K.S. Birdi, \textit{Adsorption and the Gibbs surface excess}. (Plenum Press, New York, 1984).

\bibitem{foo2010insights}
K.Y. Foo, B.H. Hameed (2010). Insights into the modeling of adsorption isotherm systems. \textit{Chemical engineering journal}, 156(1), 2-10.

\bibitem{adamson1967physical}
A.W. Adamson, A.P. Gast. \textit{Physical chemistry of surfaces}. ( Interscience publishers, New York, 1967).


\bibitem{chang1995adsorption}
C.H. Chang, E.I. Franses (1995). Adsorption dynamics of surfactants at the air/water interface: a critical review of mathematical models, data, and mechanisms. \textit{Colloids and Surfaces A: Physicochemical and Engineering Aspects}, 100, 1-45.

\bibitem{brunauer1938adsorption}
S. Brunauer, P.H. Emmett, E. Teller (1938). Adsorption of gases in multimolecular layers. \textit{Journal of the American chemical society}, 60(2), 309-319.

\bibitem{ehrlich1956mechanism}
G. Ehrlich (1956). The mechanism of chemisorption on metals. \textit{Journal of Physics and Chemistry of Solids}, 1(1-2), 3-13.

\bibitem{kisliuk1957sticking}
P. Kisliuk (1957). The sticking probabilities of gases chemisorbed on the surfaces of solids. \textit{Journal of Physics and Chemistry of Solids}, 3(1-2), 95-101.

\bibitem{noskov2020adsorption}
B.A. Noskov, A.G. Bykov, G. Gochev, S.Y. Lin, G. Loglio, R. Miller, O.Y. Milyaeva (2020). Adsorption layer formation in dispersions of protein aggregates. \textit{Advances in colloid and interface science}, 276, 102086.

\bibitem{ward1946time}
A.F.H. Ward, L. Tordai (1946). Time‐dependence of boundary tensions of solutions I. The role of diffusion in time‐effects. \textit{The Journal of Chemical Physics}, 14(7), 453-461.

\bibitem{bond1937lxxvii}
W.N. Bond, H.O. Puls (1937). LXXVII. The change of surface tension with time. \textit{The London, Edinburgh, and Dublin Philosophical Magazine and Journal of Science}, 24(164), 864-888.

\bibitem{langmuir1937effect}
I. Langmuir, V. Schaefer (1937). The effect of dissolved salts on insoluble monolayers. \textit{Journal of the American Chemical Society}, 59(11), 2400-2414.

\bibitem{ross1945change}
S. Ross (1945). The change of surface tension with time. I. Theories of diffusion to the surface. \textit{Journal of the American Chemical Society}, 67(6), 990-994.

\bibitem{carslaw1921introduction}
H.S. Carslaw. \textit{Introduction to the Mathematical Theory of the Conduction of Heat in Solids}. (MacMillan and Company, London, 1921). 


\bibitem{ward1952time}
A.F.H. Ward, L. Tordai (1952). Time‐dependence of boundary tensions of solutions: IV. Kinetics of adsorption at liquid‐liquid interfaces. \textit{Recueil des Travaux Chimiques des Pays‐Bas}, 71(6), 572-584.

\bibitem{diamant1996kinetics}
H. Diamant, D. Andelman (1996). Kinetics of Surfactant Adsorption at Fluid-Fluid Interfaces. \textit{The Journal of Physical Chemistry}, 100(32), 13732-13742.

\bibitem{miller2017dynamic}
R. Miller, E.V. Aksenenko, V.B. Fainerman (2017). Dynamic interfacial tension of surfactant solutions. \textit{Advances in colloid and interface science}, 247, 115-129.

\bibitem{sutherland1952kinetics}
K.L. Sutherland. (1952). The kinetics of adsorption at liquid surfaces. \textit{Australian Journal of Chemistry}, 5(4), 683-696.

\bibitem{delahay1957adsorption}
P. Delahay, I. Trachtenberg (1957). Adsorption kinetics and electrode processes. \textit{Journal of the American Chemical Society}, 79(10), 2355-2362.

\bibitem{hansen1961diffusion}
R.S. Hansen (1961). Diffusion and the kinetics of adsorption of aliphatic acids and alcohols at the water-air interface. \textit{Journal of Colloid Science}, 16(6), 549-560.

\bibitem{delahay1958adsorption}
P. Delahay, C.T. Fike (1958). Adsorption Kinetics with diffusion control—the plane and the expanding sphere. \textit{Journal of the American Chemical Society}, 80(11), 2628-2630.


\bibitem{miller1981solution}
R. Miller (1981). On the solution of diffusion controlled adsorption kinetics for any adsorption isotherms. \textit{Colloid and polymer science}, 259(3), 375-381.

\bibitem{mccoy1983analytical}
B.J. McCoy (1983). Analytical solutions for diffusion-controlled adsorption kinetics with non-linear adsorption isotherms. \textit{Colloid and polymer science}, 261(6), 535-539.

\bibitem{mysels1982diffusion}
K.J. Mysels (1982). Diffusion-controlled adsorption kinetics. General solution and some applications. \textit{The Journal of Physical Chemistry}, 86(23), 4648-4651.

\bibitem{frisch1983diffusion}
H.J. Frisch, K.J. Mysels (1983). Diffusion-controlled adsorption. Concentration kinetics, ideal isotherms, and some applications. \textit{The Journal of Physical Chemistry}, 87(20), 3988-3990.

\bibitem{liu2009diffusion}
J. Liu, P. Li, C. Li, Y. Wang (2009). Diffusion-controlled adsorption kinetics of aqueous micellar solution at air/solution interface. \textit{Colloid and Polymer Science}, 287(9), 1083-1088.

\bibitem{miller1991adsorption}
R. Miller, G. Kretzschmar (1991). Adsorption kinetics of surfactants at fluid interfaces. \textit{Advances in colloid and interface science}, 37(1-2), 97-121.

\bibitem{hansen1960theory}
R.S. Hansen (1960). The theory of diffusion controlled absorption kinetics with accompanying evaporation. \textit{The Journal of Physical Chemistry}, 64(5), 637-641.

\bibitem{ziller1986solution}
M. Ziller, R. Miller (1986). On the solution of diffusion controlled adsorption kinetics by means of orthogonal collocation. \textit{Colloid and polymer science}, 264(7), 611-615.

\bibitem{nguyen2006effect}
A.V. Nguyen, C.M. Phan, G.M. Evans (2006). Effect of the bubble size on the dynamic adsorption of frothers and collectors in flotation. \textit{International Journal of Mineral Processing}, 79(1), 18-26.

\bibitem{li2010simple}
X. Li, R. Shaw, G.M. Evans, P. Stevenson (2010). A simple numerical solution to the Ward–Tordai equation for the adsorption of non-ionic surfactants. \textit{Computers \& chemical engineering}, 34(2), 146-153.

\bibitem{fernandez2011numerical}
J.R. Fernández, M.C. Muñiz (2011). Numerical analysis of surfactant dynamics at air-water interface using the Henry isotherm. \textit{Journal of mathematical chemistry}, 49(8), 1624-1645.


\bibitem{liggieri1996diffusion}
L. Liggieri, F. Ravera, A. Passerone (1996). A diffusion-based approach to mixed adsorption kinetics. \textit{Colloids and surfaces A: physicochemical and engineering aspects}, 114, 351-359.

\bibitem{borwankar1983kinetics}
R.P. Borwankar, D.T. Wasan (1983). The kinetics of adsorption of surface active agents at gas-liquid surfaces. \textit{Chemical Engineering Science}, 38(10), 1637-1649.

\bibitem{adamczyk1987nonequilibrium}
Z. Adamczyk (1987). Nonequilibrium surface tension for mixed adsorption kinetics. \textit{Journal of colloid and interface science}, 120(2), 477-485.

\bibitem{miura2015diffusion}
T. Miura, K. Seki (2015). Diffusion influenced adsorption kinetics. \textit{The Journal of Physical Chemistry B}, 119(34), 10954-10961.

\bibitem{carslaw1959conduction}
H.S. Carslaw, J.C. Jaeger. \textit{Conduction of heat in solids}. (Oxford University Press, 1959). See chapter XIV.

\bibitem{redner2001guide}
S. Redner, \textit{A guide to first-passage processes}. (Cambridge
University Press, Cambridge, England, 2001).

\bibitem{klafter2011first}
J. Klafter, I.M. Sokolov. \textit{First steps in random walks: from tools to applications}, (Oxford University Press,
New York, 2011).

\bibitem{metzler2014first}
R. Metzler, S. Redner, G. Oshanin. \textit{First-passage phenomena and their applications} (World Scientific,
Singapore, 2014), Vol. 35.

\bibitem{agmon1984diffusion}
N. Agmon (1984). Diffusion with back reaction. \textit{The Journal of chemical physics}, 81(6), 2811-2817.

\bibitem{agmon1988geminate}
N. Agmon, E. Pines, D. Huppert. (1988). Geminate recombination in proton‐transfer reactions. II. Comparison of diffusional and kinetic schemes. \textit{The Journal of chemical physics}, 88(9), 5631-5638.

\bibitem{grebenkov2013geometrical}
D.S. Grebenkov, B.T. Nguyen (2013). Geometrical structure of Laplacian eigenfunctions. \textit{siam REVIEW}, 55(4), 601-667.

\bibitem{kim1999exact}
H. Kim, K.J. Shin (1999). Exact solution of the reversible diffusion-influenced reaction for an isolated pair in three dimensions. \textit{Physical review letters}, 82(7), 1578.

\bibitem{spiegel1965schaum}
M.R. Spiegel. \textit{Schaum's outline of theory and problems of Laplace transforms} (McGraw-Hill, 1965).

\bibitem{collins1949diffusion}
F.C. Collins, G.E. Kimball (1949). Diffusion-controlled reaction rates. \textit{Journal of colloid science}, 4(4), 425-437.

\bibitem{sano1979partially}
H. Sano, M. Tachiya (1979). Partially diffusion‐controlled recombination. \textit{The Journal of Chemical Physics}, 71(3), 1276-1282.

\bibitem{szabo1984localized}
A. Szabo, G. Lamm, G.H. Weiss (1984). Localized partial traps in diffusion processes and random walks. \textit{Journal of statistical physics}, 34(1-2), 225-238.

\bibitem{weiss1986overview}
G.H. Weiss (1986). Overview of theoretical models for reaction rates. \textit{Journal of Statistical Physics}, 42(1), 3-36.

\bibitem{singer2008partially}
A. Singer, Z. Schuss, A. Osipov, D. Holcman (2008). Partially reflected diffusion. \textit{SIAM Journal on Applied Mathematics}, 68(3), 844-868.

\bibitem{palreactive}
A. Pal, I.P. Castillo, A. Kundu (2019).
Motion of a Brownian particle in the presence of reactive boundaries. \textit{Physical Review E}, 100, 042128.

\bibitem{grebenkov2020imperfect}
D.S. Grebenkov (2020). Imperfect diffusion-controlled reactions. In Chemical Kinetics: Beyond the Textbook (pp. 191-219).

\bibitem{agmon1989theory}
N. Agmon, G. Weiss (1989). Theory of non‐Markovian reversible dissociation reactions. \textit{The Journal of chemical physics}, 91(11), 6937-6942.

\bibitem{abate2004multi}
J. Abate, P.P. Valkó (2004). Multi‐precision Laplace transform inversion. \textit{International Journal for Numerical Methods in Engineering}, 60(5), 979-993.

\bibitem{grebenkov2019spectral}
D.S. Grebenkov (2019). Spectral theory of imperfect diffusion-controlled reactions on heterogeneous catalytic surfaces. \textit{The Journal of Chemical Physics}, 151(10), 104108.

\bibitem{grebenkov2020paradigm}
D.S. Grebenkov (2020). Paradigm shift in diffusion-mediated surface phenomena. \textit{Physical Review Letters}, 125(7), 078102.

\bibitem{bray2000random}
A.J. Bray (2000). Random walks in logarithmic and power-law potentials, nonuniversal persistence, and vortex dynamics in the two-dimensional XY model. \textit{Physical Review E}, 62(1), 103.

\bibitem{martin2011first}
E. Martin, U. Behn, G. Germano (2011). First-passage and first-exit times of a Bessel-like stochastic process. \textit{Physical Review E}, 83(5), 051115.

\bibitem{ryabov2015brownian}
A. Ryabov, E. Berestneva, V. Holubec (2015). Brownian motion in time-dependent logarithmic potential: exact results for dynamics and first-passage properties. \textit{The Journal of chemical physics}, 143(11), 114117.

\bibitem{ray2020diffusion}
S. Ray, S. Reuveni (2020). Diffusion with resetting in a logarithmic potential. \textit{The Journal of chemical physics}, 152(23), 234110.

\bibitem{lin1990diffusion}
S.Y. Lin, K. McKeigue, C. Maldarelli (1990). Diffusion‐controlled surfactant adsorption studied by pendant drop digitization. \textit{AIChE Journal}, 36(12), 1785-1795.

\bibitem{liu2000diffusion}
J. Liu, U. Messow (2000). Diffusion-controlled adsorption kinetics at the air/solution interface. \textit{Colloid and Polymer Science}, 278(2), 124-129.

\bibitem{wang2014dependence}
C. Wang, H. Morgner (2014). The dependence of surface tension on surface properties of ionic surfactant solution and the effects of counter-ions therein. \textit{Physical Chemistry Chemical Physics}, 16(42), 23386-23393.

\bibitem{agmon1990theory}
N. Agmon, A. Szabo (1990). Theory of reversible diffusion‐influenced reactions. \textit{The Journal of Chemical Physics}, 92(9), 5270-5284.

\bibitem{agmon1993competitive}
N. Agmon (1993). Competitive and noncompetitive reversible binding processes. \textit{Physical Review E}, 47(4), 2415.

\bibitem{gopich1999excited}
I.V. Gopich, K. Solntsev, N. Agmon (1999). Excited-state reversible geminate reaction. I. Two different lifetimes. \textit{The Journal of chemical physics}, 110(4), 2164-2174.

\bibitem{prustel2012exact}
T. Prüstel, M. Meier-Schellersheim (2012). Exact Green's function of the reversible diffusion-influenced reaction for an isolated pair in two dimensions. \textit{The Journal of chemical physics}, 137(5), 054104.

\bibitem{house2007principles}
J.E. House. \textit{Principles of chemical kinetics} (Academic press, 2007), pp.145-147.

\bibitem{reva2021first}
M. Reva, D.A. DiGregorio, D.S. Grebenkov (2021). A first-passage approach to diffusion-influenced reversible binding and its insights into nanoscale signaling at the presynapse. \textit{Scientific reports}, 11(1), 1-17.

\bibitem{prustel2013reversible}
T. Prüstel, M. Tachiya (2013). Reversible diffusion-influenced reactions of an isolated pair on some two dimensional surfaces. \textit{The Journal of Chemical Physics}, 139(19), 194103.

\bibitem{grebenkov2019reversible}
D.S. Grebenkov (2019). Reversible reactions controlled by surface diffusion on a sphere. \textit{The Journal of chemical physics}, 151(15), 154103.

\bibitem{grebenkov2017first}
D.S. Grebenkov (2017). First passage times for multiple particles with reversible target-binding kinetics. \textit{The Journal of chemical physics}, 147(13), 134112.

\bibitem{lawley2019first}
S.D. Lawley, J.B. Madrid (2019). First passage time distribution of multiple impatient particles with reversible binding. \textit{The Journal of chemical physics}, 150(21), 214113.

\bibitem{grebenkov2021reversible}
D.S. Grebenkov, A. Kumar (2021). Reversible Target-Binding Kinetics of Multiple Impatient Particles. \textit{The Journal of Chemical Physics}, 156(8), 084107.

\bibitem{grebenkov2022first}
D. Grebenkov, A. Kumar (2022). First-passage times of multiple diﬀusing particles with reversible target-binding kinetics. \textit{Journal of Physics A: Mathematical and Theoretical}.

\bibitem{paul1999stochastic}
W. Paul, J. Baschnagel.  \textit{Stochastic processes. From Physics to finance} (Springer, 1999). 

\end{thebibliography}
\end{document}